\documentclass[epj]{svjour}

\usepackage{amsmath} 
\usepackage{amssymb} 
\usepackage{graphicx} 
\usepackage{color}

\newcommand{\bsigma}{\mbox{\boldmath $\sigma$}} 
\newcommand{\bsigmaa}{\mbox{\boldmath $\sigma'$}} 
\newcommand{\btau}{\mbox{\boldmath $\tau$}}
\newcommand{\nlim}{n\rightarrow 0}
\newcommand{\Nlim}{N\rightarrow\infty}

\newcommand{\pf}{\mathcal{Z}}
\newcommand{\trace}{\mbox{Tr}}
\newcommand{\bPsi}{{\bf \Psi}}

\newcommand{\bPhi}{{\bf \Phi}}

\begin{document}

\authorrunning{ D. Boll\'e, J. Busquets Blanco}
\titlerunning{The synchronous BEG neural network with variable dilution}

\title{The Blume-Emery-Griffiths neural network with synchronous updating and variable dilution }
\author{D. Boll\'e \and J. Busquets Blanco 
\thanks{\emph{e-mail:} Jordi.Busquets@fys.kuleuven.be}%
}                     
%
%
\institute{Instituut voor Theoretische Fysica, 
                 Katholieke Universiteit Leuven, 
                 Celestijnenlaan 200 D, 
                 B-3001 Leuven, Belgium }
\date{Received: date / Revised version: date}
%
\abstract{The thermodynamic and retrieval properties of the Blume-Emery-Griffiths
neural network with synchronous updating and variable dilution are studied
using replica mean-field theory. Several forms of dilution are allowed 
by pruning the different types of couplings present in the Hamiltonian. 
The appearance and properties of two-cycles are discussed. Capacity-temperature phase diagrams are derived for several values of the pattern activity. The results are compared with those for sequential updating. The effect of self-coupling is studied. Furthermore, the optimal combination of dilution parameters giving the largest critical capacity is obtained.  
%
} 
\maketitle

\section{Introduction} \label{sec:intro}

Recently, we have studied the possible different physics arising from sequential and synchronous updating of the spins in multi-state Ising-type ferromagnets \cite{BB04}. In general, it is known that besides fixed-point attractors, synchronous updating in  spin-models can give rise to cycles as stationary behaviour \cite{Pbook,C01s}. In \cite{BB04} we have shown that for the $Q=3$ Ising ferromagnet with synchronous updating the same stationary solutions appear as for sequential updating except for negative couplings. In that case, for synchronous updating,  cycles of period two in the magnetization occur in the ferromagnetic phase. For the  Blume-Emery-Griffiths (BEG) ferromagnet, however, we have found that synchronous updating allows for a much richer free energy landscape and phase diagram. 

In view of these results it is of interest to extend such a study to multi-state neural networks in order to see  whether the different types of updating give rise to different retrieval properties. For the Hopfield neural network model, e.g., involving binary neurons, it is known that in a replica-symmetric treatment the retrieval region is slightly larger in the synchronous case \cite{FK88}. A natural question is then whether a similar effect is present in the BEG neural network, which is known to optimize the mutual information content for three-state networks \cite{DK00,BV02} and, hence, to guarantee the best retrieval properties in comparison with other three-state networks (see \cite{bollebook} for a review). For sequential updating, the thermodynamic and retrieval properties are studied in detail in the literature for fully connected architectures  \cite{BV03} and diluted architectures \cite{BDEKT03,tonithe} (and references therein), for synchronous updating such a study has not yet appeared.

The purpose of this work is precisely to report on a study of the BEG neural network with synchronous updating and variable dilution. In particular, we discuss the thermodynamic and retrieval properties of this model using replica symmetric mean-field theory. Several forms of dilution are allowed by cutting some of the couplings appearing in the Hamiltonian: the couplings multiplying the term bilinear in the neurons and/or the couplings related to the term biquadratic in the neurons. 
Capacity-temperature phase diagrams are derived for several values of the pattern activity. 
In  particular, the appearance and properties of cycles are discussed. Apart from these cycles it turns out that the asymptotic behaviour is almost identical to the one for sequential updating. For some values of the (fully-connected) network parameters the retrieval region is marginally enhanced. Only the addition of self-coupling, not allowed in the sequential model if one wants to satisfy detailed balance  \cite{C01s,Peretto}, can enlarge the retrieval region substantially, especially in the case of dilution. The optimal combination of the network dilution parameters giving the largest critical capacity is obtained.

The rest of the paper is organized as follows. In Section \ref{2} the model with variable dilution is introduced from a dynamic point of view. Section \ref{3} presents the replica-symmetric mean-field calculation of the free energy and obtains the fixed-point equations for the relevant order parameters. The possible appearance and properties of two-cycles are examined. In Section \ref{4} these equations are studied in detail for arbitrary temperatures and different types of dilution. In particular, capacity-temperature phase diagrams are obtained and compared with the corresponding results for sequential updating. The optimal dilution parameters leading to the largest critical capacity are obtained.  Also the specific thermodynamic properties are discussed.  Furthermore, the effect of self-coupling on the retrieval properties and on the phase-diagrams is analized. 
Some concluding remarks are given in Section \ref{6}.

\section{The model} 
\label{2}

Consider a network of N neurons, $\bsigma=\{\sigma_1,...,\sigma_N\}$,
which can take
values from the set $\mathcal{S}=\{-1, 0, 1\}$. In this network we want to store
$p=\alpha N$ patterns, $\{\xi_i^\mu\}$, $i = 1, \dots, N$ and $\mu = 1, \dots, p$.
They are supposed to be independent identically distributed random variables
(iidrv) with respect to $i$ and $\mu$, drawn from a probability distribution
given by
\begin{equation}
\textrm{Pr}(\xi_i^\mu) = a \, \delta(1 - (\xi_i^\mu)^2)
                     + (1-a) \, \delta(\xi^\mu_i)\, ,
\end{equation}
with $a$  the pattern activity, namely  
\begin{equation}
\textrm{E}((\xi_i^\mu)^2) = a \ .
\end{equation}

The neurons are updated synchronously according to the transition probability 
\begin{eqnarray}
&&\mbox{Pr}\left(\bsigma|\bsigmaa\right)=\prod_{i=1}^N\mbox{Pr}\left(\sigma_i=s\in \mathcal{S}|\bsigmaa \right) \\
\label{three}
&&\mbox{Pr}\left(\sigma_i=s\in\mathcal{S}|\bsigmaa \right)=
     \frac{\exp[-\beta \epsilon_i(s|\bsigmaa)]}
          {\displaystyle{\sum_{s \in\mathcal{S}} 
                    \exp[-\beta \epsilon_i(s|\bsigmaa)] }}
\end{eqnarray}
with $\beta$ the inverse temperature and $\epsilon_i(s|\bsigma)$ an
effective single site energy function given by
\begin{equation}
\label{four}
\epsilon_i(s|\bsigma)= 
  -h_{1,i}(\bsigma)s -h_{2,i}(\bsigma) s^2 \, ,
           \quad s \in \mathcal{S}\, ,
\end{equation}
where the random local fields are defined by
\begin{equation}
h_{k,i}(\bsigma)=\sum_{j=1}^N J^c_{k,ij} \sigma_j^k \, , \quad k=1,2 \, .
\end{equation}
In the sequel the index $k$ will always take the values $k=1,2$.
The couplings $J^c_{k,ij}$ are taken to be of the form 
\begin{equation}
J_{k,ij}^c=\frac{c_{k,ij}}{c_k}J_{k,ij}
\label{coupdil}
\end{equation}
where the probability distribution of the $\{c_{k,ij}\}$ is given by
\begin{equation}
\mbox{Pr}(c_{k,ij})=c_k\delta(c_{k,ij}-1)+(1-c_k)\delta(c_{k,ij})\,.
\label{propdil}
\end{equation}
Hence, they allow for a diluted architecture. The  $J_{k,ij}$ are determined
via the Hebb rule
\begin{equation}
J_{k,ij} = \frac{1}{a_k^2 N} \sum_{\mu=1}^p \eta^\mu_{k,i} \eta^\mu_{k,j} 
\label{couphebb}
\end{equation}
where $\eta_{1,i}^\mu=\xi_{i}^\mu$ and $\eta_{2,i}^\mu=((\xi_{i}^\mu)^2-a)$.
For convenience we define $a_1\equiv
\textrm{E}((\eta_{1,i}^\mu)^2)=a$, $a_2\equiv\textrm{E}((\eta_{2,i}^\mu)^2)=a(1-a)$.

As can be seen from  (\ref{coupdil})-(\ref{couphebb}), two types of dilution are 
possible. Some of the couplings related to the bilinear term in the neurons, $k=1$,
can be cut and/or some of the couplings related to the biquadratic term, $k=2$, can be 
pruned, both simultaneously or independently. In the following we denote this as $k$-type dilution. We remark that in the literature $k=2$-type dilution is frequently indicated as dilution of the active patterns, since only the fact whether the neuron is active or not (and not the sign of the neuron state) is important.
In order for the model to satisfy detailed balance, the dilution has to be 
symmetric ($c_{k,ij}=c_{k,ji}$). In the case of extreme dilution, when  $c_k=0$, 
the average number of connections per neuron, $c_k N$, is still taken to be 
infinite, as usual. This can be achieved by requiring that 
\begin{equation}
\lim_{c_{k}\rightarrow 0}c_kN=\log{N}\, .
\end{equation}

Finally, we recall that the detailed balance property for synchronous updating
is not destroyed by the presence of self-couplings, i.e., couplings of the form 
$J_{k,ii}^c$, $k=1,2$. Hence, we do allow for this type of couplings
and write them as  $J_{k,ii}^c=J_{k,0}\alpha_k/a_k$, with
$J_{k,0}$ a parameter and $\alpha_k=p/c_k N$ the $k$-type capacities. 

The long-time behaviour of this model is governed by the pseudo-Hamiltonian 
\cite{BB04}
\begin{equation}
H(\bsigma)=-\frac{1}{\beta}\sum_{i=1}^{N}\ln{\left\{2
    \exp{(\beta h_{2,i}(\bsigma))}
    \cosh{(\beta h_{1,i}(\bsigma))}+1\right\}}
\end{equation}
with as duplicate-neuron (spin) representation
\begin{equation}
H(\bsigma, \btau)=-\sum_{k}\sum_{i,j}J_{k,ij}^c\sigma_i^k\tau_j^k
\label{hamtwospin}
\end{equation}
such that $\underset{\bsigma}{\trace}\,[\exp(- \beta H(\bsigma))]= 
\underset{\bsigma}{\trace}\underset{\btau}{\trace}\,[\exp(- \beta  H(\bsigma, \btau))]$. 

In the next Sections we study the thermodynamic and retrieval properties of this 
model starting from the free energy.

\section{Replica mean-field theory}
\label{3}

\subsection{Replicated free energy}

We adapt the standard replica technique as applied to dilute models \cite{WS91}-\cite{TE01} to synchronous updating in order to calculate 
the free energy of the model. Since the method is rather standard by now, although the specific details are rather involved, we only indicate the main steps. For the completely detailed calculation we refer to  \cite{jordithe}. 

We start from the replicated partition function based upon the two-spin Hamiltonian (\ref{hamtwospin})
\begin{eqnarray}
\left\langle \pf^n\right\rangle_{c,\xi}
=\prod_\rho\left(
\underset{\bsigma_\rho}{\trace}
\underset{\btau_\rho}{\trace}
\right)
\prod_{k,i}
      \exp{\left(\beta\frac{\alpha_k}{a_k}J_{k,0}\sum_{\rho}
           \sigma_{i,\rho}^k\tau_{i,\rho}^k\right)} 
	   \nonumber \\
\left\langle\prod_k\prod_{i,j\neq i}\exp{\left(
     \beta J_{k,ij}^c\sum_{\rho}\sigma_{i,\rho}^k\tau_{i,\rho}^k
\right)}\right\rangle_{c,\xi}
\end{eqnarray}
where the average is over the patterns and the dilution and where we have explicitly separated off the self-interactions.

We first average over the dilution by expanding the exponential with respect to 
$J_{k,ij}^c$ up to order $\mathcal{O}((c_kN)^{-3/2})$. After using that the couplings $J_{k,ij}\sim\mathcal{O}((\alpha_k c_k/a_k^2 N)^{1/2})$ and assuming a macroscopic overlap with only one pattern we arrive at 
\begin{eqnarray}
&&\left\langle \pf^n\right\rangle_{c,\xi}
=
\prod_\rho\left(
\underset{\bsigma_\rho}{\trace}
\underset{\btau_\rho}{\trace}
\right)
\,\mathcal{Z}^c\langle\mathcal{Z}^{nc}\rangle_\xi
         \nonumber\\
&& \times \prod_k\Bigg\{
\exp{\Big(\frac{\beta\alpha_k}{a_k}(J_{k,0}-c_k)\sum_{i,\rho}
           \sigma_{i,\rho}^k\tau_{i,\rho}^k\Big)}
	      \nonumber\\
&&\times\exp{\Big(
\frac{\beta^2\alpha_k(1-c_k)}{4a_k^2N}\sum_{i,j\neq i}\Big[\sum_\rho
\Big(\sigma_{i,\rho}^k\tau_{j,\rho}^k+
\sigma_{j,\rho}^k\tau_{i,\rho}^k\Big)\Big]^2\Big)}\Bigg\}\,  
\nonumber \\
\end{eqnarray}
where $\pf^c$ denotes the condensed part (e.g., pattern $\mu=1$) and $\pf^{nc}$ the non-condensed part (patterns $\mu=2,...,p$) 
\begin{equation}
\mathcal{Z}^c\langle\mathcal{Z}^{nc}\rangle_\xi
=
\left\langle
     \prod_k\Bigg(\prod_{i,j}\exp{\Big(\beta \sum_{\rho}
     J_{k,ij}\sigma_{i,\rho}^k\tau_{j,\rho}^k\Big)} \Bigg)
     \right\rangle_\xi\,.
\end{equation}

Next, we consider the average over the disorder. A Hubbard-Stratonovich 
transformation allows us to write the condensed part, introducing 
$\sqrt{N \beta}m_{kk',\rho}^1$ with $k,k'=1,2$  (we forget about the upperindex $1$ in the sequel), as 
\begin{eqnarray}
&& \pf^c \varpropto \int \prod_{k,k'}\Bigg(d{\bf m}_{kk'}\,
       \exp{\Bigg(-\frac{\beta N}{2}\sum_\rho m_{kk',\rho}^2\Bigg)}\Bigg)
       \nonumber \\
&&     \times\prod_k\exp \left(\frac{\beta}{a_k}\sum_{\rho}\sum_i\left[
          \eta_{k,i}(\sigma_{i,\rho}^k+\tau_{i,\rho}^k)m_{k1,\rho}  \right.\right. 
	  \nonumber \\
&&     \hspace{2cm} \left.\left. + 
           i\eta_{k,i}(\sigma_{i,\rho}^k-\tau_{i,\rho}^k)m_{k2,\rho}
              \right] \rule{0cm}{0.6cm} \right)
	      \label{avcon}   
\end{eqnarray}
with $d{\bf m}_{kk'}=\prod_{\rho}m_{kk',\rho}$.  For the non-condensed part
we first do a Hubbard-Stratonovich transformation introducing the variables  
$m_{kk',\rho}^\mu$ for $\mu>1$, expand then the exponential up to order 
$\mathcal{O}(N^{-3/2})$, and finally use $\cosh(x)\simeq \exp(x^2/2)$ to obtain
\begin{equation}
\langle\pf^{nc}\rangle_\xi^{1/(1-p)}\varpropto\int d\bPhi\,\exp{\left(-\frac{1}{2}
    {\bf \Phi}^2+\frac{\beta}{2}{\bf \Phi}\bPsi{\bf \Phi}^{\dagger}\right)}\, ,
\label{addition}
\end{equation}
where ${\bf \Phi}$ is a $4n$-dimensional vector of the form
${\bf \Phi}=(\bPhi_1, \bPhi_2, ..., \bPhi_n)$ with
$\bPhi_{\rho}=(m_{11,\rho},m_{12,\rho},m_{21,\rho},m_{22,\rho})$. In the
same way, $\bPsi$ is a $4n$ x $4n$ matrix of the form
\begin{displaymath}
\bPsi=
\left( \begin{array}{cccc}
\bPsi_{11} & i\bPsi_{12} & 0 & 0 \\
i\bPsi_{13} & -\bPsi_{14} & 0 & 0 \\
0 & 0 & \bPsi_{21} & i\bPsi_{22} \\
0 & 0 & i\bPsi_{23} & -\bPsi_{24} \\
\end{array} \right)\, .
\end{displaymath}
Each element in itself is an $n$ x $n$ matrix with  elements defined in terms of the neuron variables as
\begin{eqnarray}
\psi_{k1,\rho\gamma}&=&\frac{1}{2Na_k}\sum_{i=1}^N(\sigma_{i,\rho}^k+\tau_{i,\rho}^k)
                                              (\sigma_{i,\gamma}^k+\tau_{i,\gamma}^k)
                                              \, ,\nonumber\\
\psi_{k2,\rho\gamma}&=&\frac{1}{2Na_k}\sum_{i=1}^N(\sigma_{i,\rho}^k+\tau_{i,\rho}^k)
                                              (\sigma_{i,\gamma}^k-\tau_{i,\gamma}^k)
                                              \, ,\nonumber\\
\psi_{k3,\rho\gamma}&=&\frac{1}{2Na_k}\sum_{i=1}^N(\sigma_{i,\rho}^k-\tau_{i,\rho}^k)
                                              (\sigma_{i,\gamma}^k+\tau_{i,\gamma}^k)
                                              \, ,\nonumber\\
\psi_{k4,\rho\gamma}&=&\frac{1}{2Na_k}\sum_{i=1}^N(\sigma_{i,\rho}^k-\tau_{i,\rho}^k)
                                              (\sigma_{i,\gamma}^k-\tau_{i,\gamma}^k)
                                              \, .\nonumber
\end{eqnarray}
The integral over the $\bPhi$ in (\ref{addition}) can be done, yielding
\begin{eqnarray}
&& \langle\pf^{nc}\rangle_\xi\varpropto\exp{\left(-\frac{N}{2}\sum_{k}\alpha_kc_k
        \trace\big(\ln{\left[{\bf I}- \beta\bPsi_{k1}\right]}\big)\right)}\nonumber\\
&&  \times\exp\left(-\frac{N}{2}\sum_{k}\alpha_kc_k\trace\big(\ln
   \left[{\bf  I}+ \beta \bPsi_{k4}         \right. \right. \nonumber \\
&& \left. \left. \hspace*{2cm} + \beta^2\bPsi_{k3}({\bf I}-\beta\bPsi_{k1})^{-1}\bPsi_{k2}
		\right]\big)\rule{0cm}{0.6cm}\right)\,.
\end{eqnarray}

At this point we introduce the relevant Edwards- \\ Anderson (EA) order parameters
\begin{eqnarray}
&& q_{k1,\rho\gamma}=\frac{1}{N}\sum_{i=1}^N\sigma_{i,\rho}^k\sigma_{i,\gamma}^k
\label{EAorder0} \\
&& q_{k2,\rho\gamma}=\frac{1}{N}\sum_{i=1}^N\tau_{i,\rho}^k\tau_{i,\gamma}^k
        \label{EAorder1} \\
&& r_{k,\rho\gamma}=\frac{1}{N}\sum_{i=1}^N\sigma_{i,\rho}^k\tau_{i,\gamma}^k 
         \label{EAorder2}
\end{eqnarray}
and we redefine the $m_{kk',\rho}$  for technical reasons (recall, e.g.,  eq.(\ref{avcon}))
such that the $\tau$ and the $\sigma$ can be considered separately
\begin{equation}
m_{kk',\rho}= m_{k1,\rho}+(-1)^{k'-1}im_{k2,\rho}\, .
\end{equation}

Using its self-averaging property in the thermodynamic limit, the replicated free energy can then be written in terms of these redefined overlaps, the EA order parameters and their
conjugate variables as
\begin{eqnarray}
f_n&=& \lim_{\Nlim} -\frac{1}{N \beta}
    \log \left(\langle \pf^n\rangle_{c,\xi}\right)
\nonumber \\
 &=&\sum_{k}\sum_\rho m_{k1,\rho}m_{k2,\rho}
              -\sum_{k}\frac{\alpha_k}{a_k}(J_{k,0}-c_k)\sum_{\rho}r_{k,\rho\rho}
	      \nonumber\\
&+&\frac{1}{2\beta}\sum_k\alpha_kc_k
               \trace\left(\ln{\left[{\bf \Sigma}_k\right]}\right)\nonumber\\
&-&\frac{\beta}{2}\sum_k\alpha_k\sum_{\rho,\gamma}
                                 \Big(2r_{k,\rho\gamma}\tilde{r}_{k,\rho\gamma}+\sum_{k'}
                                        q_{kk',\rho\gamma}\tilde{q}_{kk',\rho\gamma}\Big)
     \nonumber\\
&-&\frac{\beta}{2}\sum_k\frac{\alpha_k(1-c_k)}{a_k^2}\sum_{\rho,\gamma}
              \Big(q_{k1,\rho\gamma}q_{k2,\rho\gamma}+r_{k,\rho\gamma}r_{k,\gamma\rho}\Big)
 \nonumber\\
&-&\frac{1}{\beta}\left\langle\log{\left[
\prod_{\rho} \left( \underset{\sigma_\rho}{\trace}\underset{\tau_\rho}{\trace}\right)
\exp{\left(\beta \tilde{H}(\bar{\sigma},\bar{\tau})\right)}\right]}\right\rangle_{\xi^c}
\end{eqnarray}
with $\bar{\sigma}=\{\sigma_1, \ldots, \sigma_n\}$ denoting a vector in replica space and
\begin{eqnarray}
{\bf \Sigma}_k &=& ({\bf I}-\beta\bPsi_{k1})({\bf I}+\beta\bPsi_{k4})
   \nonumber \\ 
                 &+&  \beta^2({\bf I}-\beta\bPsi_{k1})\bPsi_{k3}
                   ({\bf I}-\beta\bPsi_{k1})^{-1}\bPsi_{k2} \, .
\end{eqnarray}   
Here $\tilde{H}(\bar{\sigma},\bar{\tau})$ is the effective single-site replicated Hamiltonian
\begin{eqnarray}
&&\tilde{H}(\bar{\sigma},\bar{\tau})=\sum_k\left(\frac{1}{a_k}\eta_k\sum_{\rho}
                       (\sigma_{\rho}^km_{k1,\rho}+\tau_{\rho}^km_{k2,\rho})\right.
		       \nonumber\\
&&\left.+\frac{\beta\alpha_k}{2}\sum_{\rho,\gamma}
                 \left(2\tilde{r}_{k,\rho\gamma}\sigma_{\rho}^k\tau_{\gamma}^k+
                  \tilde{q}_{k1,\rho\gamma}\sigma_{\rho}^k\sigma_{\gamma}^k+
         \tilde{q}_{k2,\rho\gamma}\tau_{\rho}^k\tau_{\gamma}^k\right)\right)\, .
\nonumber \\
\end{eqnarray}

We remark that the $\bPsi_{kl}$ can be expressed in terms of the order parameters
defined in (\ref{EAorder0})-(\ref{EAorder2})
\begin{eqnarray}
&&{\bf \Psi}_{k1}=\frac{1}{2a_k}\big({\bf q}_{k1}+{\bf r}_{k}
           +{\bf r}_{k}^{\dagger}+{\bf q}_{k2}\big)
        \\
&&{\bf \Psi}_{k2}=\frac{1}{2a_k}\big({\bf q}_{k1}-{\bf r}_{k}
         +{\bf r}_{k}^{\dagger}-{\bf q}_{k2}\big)
     \\
&&{\bf \Psi}_{k3}=\frac{1}{2a_k}\big({\bf q}_{k1}+{\bf r}_{k}
      -{\bf r}_{k}^{\dagger}-{\bf q}_{k2}\big)
    \\
&&{\bf \Psi}_{k4}=\frac{1}{2a_k}\big({\bf q}_{k1}-{\bf r}_{k}
         -{\bf r}_{k}^{\dagger}+{\bf q}_{k2})\, .
\end{eqnarray}

The physical meaning of the parameters becomes clear when writing
\begin{eqnarray}
&&m_{k1,\rho}=\frac{1}{a_k}\left\langle \eta_k
\left\langle \sigma_\rho^k \right\rangle_{\beta}\right\rangle_{\xi^c},
q_{k1,\rho\gamma}=\left\langle \left\langle
\sigma_\rho^k\sigma_\gamma^k\right\rangle_{\beta}\right\rangle_{\xi^c}
 \\
&&m_{k2,\rho}=\frac{1}{a_k}\left\langle\eta_k
  \left\langle\tau_\rho^k\right\rangle_{\beta}\right\rangle_{\xi^c},
q_{k2,\rho\gamma}=\left\langle\left\langle
 \tau_\rho^k\tau_\gamma^k\right\rangle_{\beta}\right\rangle_{\xi^c}
  \\
&&r_{k,\rho\gamma}=\left\langle\left\langle
\sigma_\rho^k\tau_\gamma^k\right\rangle_{\beta}\right\rangle_{\xi^c},
\end{eqnarray}
where $\left\langle \cdot \right\rangle_{\xi^c}$ denotes the average over the condensed pattern and initial conditions and where the thermal average  $\left\langle \cdot \right\rangle_{\beta}$ is defined as
\begin{eqnarray}
&& \left\langle A(\bar{\sigma},\bar{\tau})\right\rangle_{\beta} = \nonumber \\
&& \hspace{1cm}  \frac 
  {\underset{\rho}{\prod}
    \left(\underset{\sigma_\rho}{\mbox{Tr}}
                       \underset{\tau_\rho}{\mbox{Tr}} \right) 
    A(\bar{\sigma},\bar{\tau})
    \exp{(\beta\tilde{H}(\bar{\sigma},\bar{\tau}))}}
       { \underset{\rho}{\prod} \left(\underset{\sigma_\rho}{\mbox{Tr}}
                    \underset{\tau_\rho}{\mbox{Tr}} \right)
       \exp{(\beta\tilde{H}(\bar{\sigma},\bar{\tau}))}}\,  
 \end{eqnarray}

Next, we want to study the replicated free energy and its minima within, as usual, the replica-symmetric approximation in order to obtain the phase diagrams for the model. To make this picture complete we also have to address the occurrence of two-cyles as possible stationary solutions of the relevant saddle-point equations.

\subsection{Replica symmetry and non-cycle ansatz}

In \cite{FK88} it has been stated, within a replica-symmetric treatment, that in the Little-Hopfield model (\cite{L74,H82}) cycles do not occur in the replica-symmetric retrieval phase. In that work the authors assume, however, that the order parameter $ r_{k,\rho\gamma}$ (recall (\ref{EAorder2})) is symmetric, implying already absence of two-cycles in the retrieval phase right from the start. 
Extensive simulations for the Little-Hopfield model have shown \cite{tonithe} that, at least in the retrieval phase, cycles do not occur. 

\begin{figure}[ht]
\begin{center}
\resizebox{0.8\columnwidth}{!}{
 \includegraphics*[angle=270,scale=0.65]{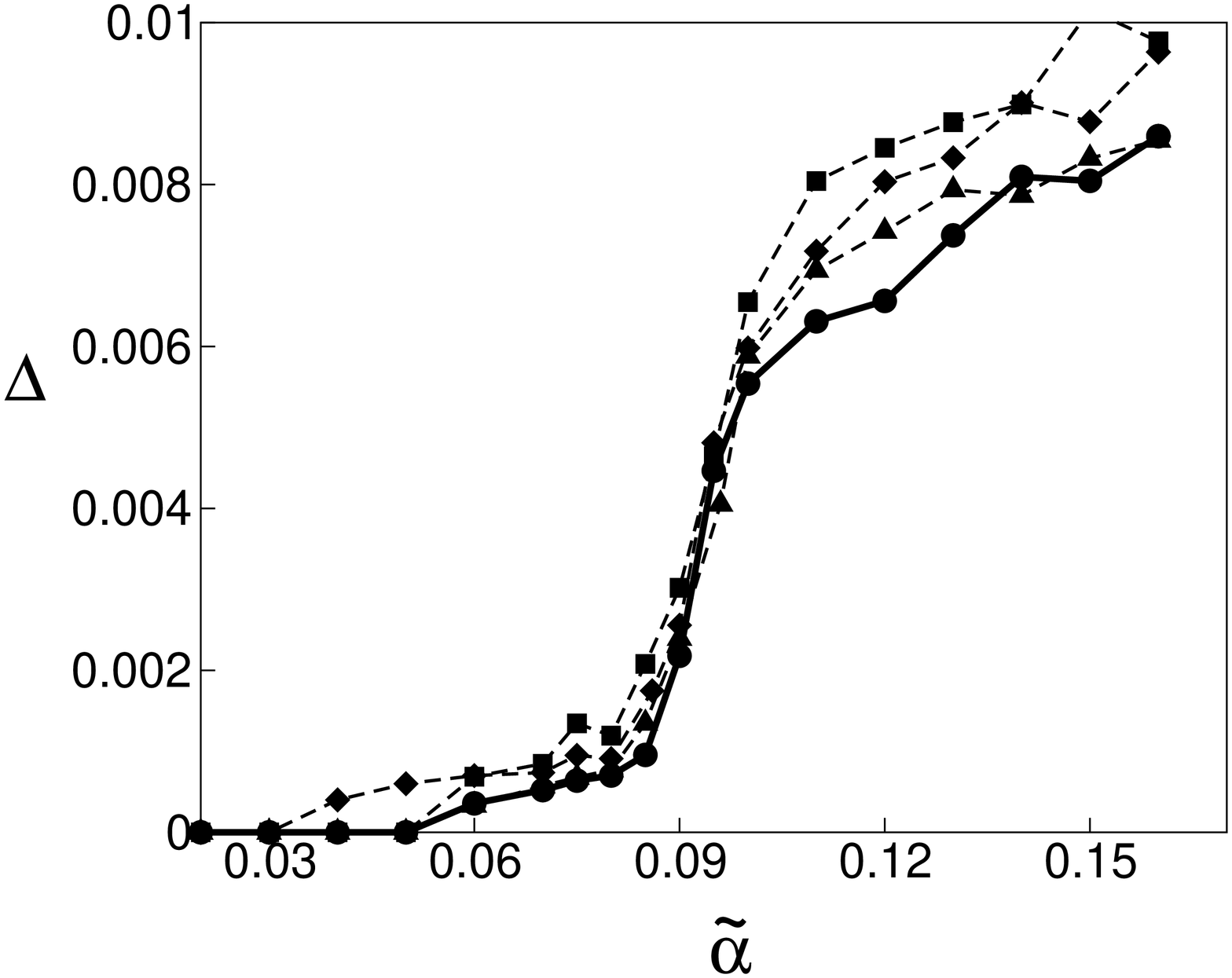}}
 \\
\resizebox{0.8\columnwidth}{!}{ 
 \includegraphics*[angle=270,scale=0.65]{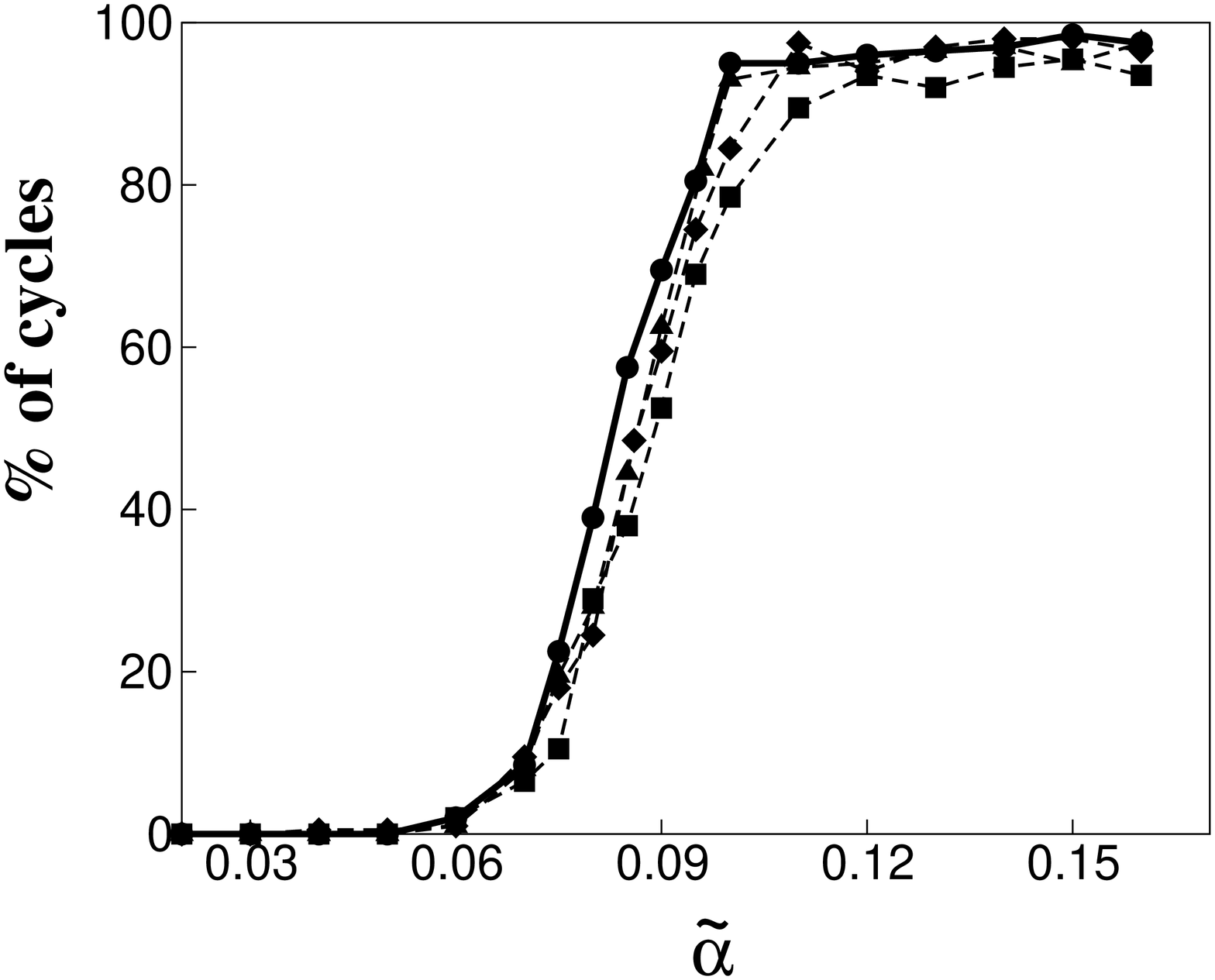}}
\\
\resizebox{0.8\columnwidth}{!}{ 
 \includegraphics*[angle=270,scale=0.65]{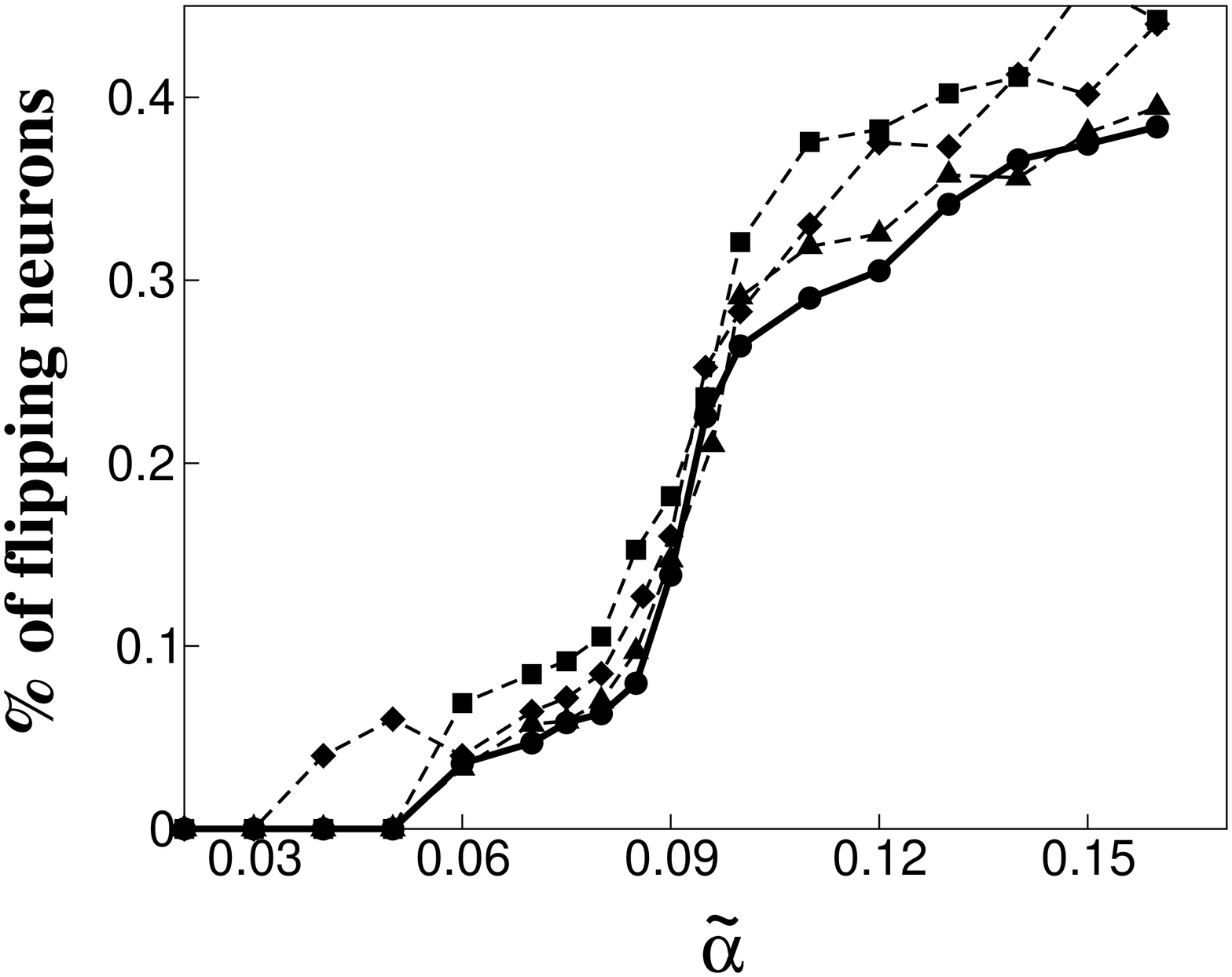}}
\caption{ Simulations with $N=4000$ (squares), $N=5000$ (diamonds), $N=6000$ (triangles) and $N=7000$ (circles) for $T=0$ and $m_k(0)=0.6, q_0(0)=0.7$. From top to bottom, we show the average value
of $\Delta(t)$ for the runs that reached a cycle, the percentage of cycles among
the 200 runs performed, and the average percentage of neurons
flipping in the cycles. Only the results for the largest system size 
have been plotted with a full line.}
   \label{fig:parstcs:cycles}       
\end{center}
\end{figure}

We have performed extensive simulations for the fully connected BEG neural network with equal dilution parameters (implying $\alpha_1=\alpha_2\equiv\tilde{\alpha}$) for uniform patterns ($a=2/3$) with synchronous updating at zero temperature. The basic quantity we have studied is the \emph{cycle parameter} $\Delta(t)$, defined as
\begin{equation}
\Delta(t)=\frac{1}{N}\sum_{i=1}^N(\sigma_i(t)-\sigma_i(t-1))^2\, .
\end{equation}
The results are shown in Fig.~\ref{fig:parstcs:cycles}. 
In particular, we have run simulations for several values of the capacity $\tilde{\alpha}$ at zero temperature during 1000 time steps of synchronous updating, which in all cases reached stationarity (i.e. a constant value $\Delta(t) \geq 0$).  
We have determined how many of the runs (200 in total) end up in a cycle and how
many in a fixed-point. Moreover, we have computed the average value of $\Delta(t)$ for the cycles. In addition, we have checked what is the average number of neurons involved in the cycles. To have an idea of the  finite size effects, the simulations are performed for $N=4000$ up to $N=7000$ neurons. From the results it seems natural to conjecture that in the limit $N\rightarrow\infty$ no cycles will exist in the retrieval region  
($\tilde{\alpha}\leqslant 0.091$ for $T=0$ \cite{BV03}). In Fig.~1 we show results for the initial conditions $m_k(0)=0.6, \, q_0(0)=0.7$ but simulations for several other initial conditions (even for $m_k(0)=0$) lead to the same conclusion.
 
The first order transition from retrieval to spin-glass behaviour includes as well 
a transition from a region where two-period cycles are not occurring to a region in which
they basically dominate the space of solutions, as can be concluded from the figure showing the percentage of cycles.

Furthermore, we see that the cycles never involve a large number of neurons 
(always less than 0.5\% of the total in the figures). The average number of flipping neurons increases monotonically with $\tilde{\alpha}$ but does seem to saturate with increasing system size. 
This is in agreement with the results obtained in a finite-size scaling study of the capacity problem for the Hopfield model \cite{SMK96} where cycles also involve only a tiny fraction of all neurons. Moreover, these authors find that there is always a fixed-point solution of the sequential updating nearby. In this sense, cycles seem to be stationary solutions in which the system is unable to reach an accessible spin-glass minimum but jumps around it.

Given these findings, we neglect cycles in the sequel of the replica calculation since we are interested mainly in the retrieval phase. 
In order to implement this non-cycle ansatz, we eliminate any distinction between $\sigma$ and $\tau$. Therefore, we introduce the following assumptions 
\begin{eqnarray}
&& m_{k1,\rho}=m_{k2,\rho}=m_{k,\rho} \\
&& q_{k1,\rho\gamma}=q_{k2,\rho\gamma}=q_{k,\rho\gamma}\, ,\quad
\tilde{q}_{k1,\gamma}=\tilde{q}_{k2,\rho\gamma}=\tilde{q}_{k,\rho\gamma}\\
&& r_{k,\rho\gamma}=q_{k,\rho\gamma}\, ,\quad\textrm{only for}\,\,\rho\neq\gamma\,.
\end{eqnarray}
We explicitly remark that introducing  $r_{k,\rho\rho}=q_{k,\rho\rho}$ would not be correct since both quantities imply products of spins from the same replica, but $\sigma$ and $\tau$ are affected  by different thermal fluctuations.

Finally, we impose symmetry in the replica space. Before doing so we solve for the conjugate
variables in order to further simplify the free energy. This allows us 
to replace $\tilde{r}_{k,\rho\gamma}$ by  $\tilde{q}_{k,\rho\gamma}$ for $\rho\neq\gamma$.  The replica symmetry (RS) ansatz then implies
\begin{eqnarray}
&&m_{k,\rho}=m_k\\
&&q_{k,\rho\gamma}=q_{k}\, ,\quad q_{k,\rho\rho}=q_{0}\\
&&\tilde{q}_{k,\rho\gamma}=\tilde{q}_{k}\, ,\quad\tilde{q}_{k,\rho\rho}=\tilde{q}_{k,0}\\
&&r_{k,\rho\rho}=r_{k,0}\, ,\quad\tilde{r}_{k,\rho\rho}=\tilde{r}_{k,0}\,.
\end{eqnarray}
As a result, the free energy per spin in the limit $\nlim$ can be written as
\begin{eqnarray}
\label{beforetracef}
&&\lim_{n\rightarrow 0}\frac{f_n}{n}= \sum_km_k^2 \nonumber\\
&& +\frac{1}{2\beta}
        \sum_k\alpha_kc_k\left(\ln{\left[(1-\chi_{kr})^2-\chi_k^2\right]}-
        \frac{2\beta q_k}{a_k(1-\chi_{kr}-\chi_k)}\right)
	\nonumber\\
  &&-\sum_k\frac{\alpha_k}{a_k}(J_{k,0}-c_k)r_{k,0}
        \nonumber \\
   &&  -\frac{\beta}{2}\sum_k\frac{\alpha_k(1-c_k)}{a_k^2}
        \left(q_0^2-2q_k^2+r_{k,0}^2\right)
	\nonumber\\
  &&+\frac{\beta}{2}\sum_k\alpha_k(
        2q_0\tilde{q}_{k,0}-4q_{k}\tilde{q}_{k}+2r_{k,0}
        \tilde{r}_{k,0})
	\nonumber\\
   &&-\frac{1}{\beta}\left\langle\int \mathcal{D}z_1\mathcal{D}z_2\log{\left[
        \underset{\sigma}{\trace}\underset{\tau}{\trace}
        \exp{\Big(\beta \tilde{H}(\sigma,\tau;z_k)\Big)}
        \right]}\right\rangle_{\xi^c}
\end{eqnarray}
with $z_1$ and $z_2$ Gaussian variables with measure $ \mathcal{D}z= dz (2 \pi)^{-1/2} \exp(-z^2/2)$. The effective Hamiltonian now reads
\begin{eqnarray}
\label{beforetraceH}
&&\tilde{H}(\sigma,\tau;z_k)=\sum_k\Big(\frac{1}{a_k}\eta_k m_k(\sigma^k+\tau^k)+
       \alpha_ka_k\tilde{\chi}_{kr}\sigma^k\tau^k\nonumber\\
 && \hspace{1cm}+\frac{1}{2}(\sigma^2+\tau^2)\alpha_ka_k\tilde{\chi}_k+
        \sqrt{\alpha_k\tilde{q}_k}(\sigma^k+\tau^k)z_k\Big)
\end{eqnarray}
with the implicit definitions for the susceptibilities 
\begin{eqnarray}
&&\chi_{k}=\frac{\beta}{a_k}(q_0-q_k)\, ,\,\,\chi_{kr}=\frac{\beta}{a_k}(r_{k,0}-q_k)\, 
 \\
&& \tilde{\chi}_{k}=\frac{\beta}{a_k}(\tilde{q}_{k,0}-\tilde{q}_k)\, ,\,\,\tilde{\chi}_{kr}=
\frac{\beta}{a_k}(\tilde{r}_{k,0}-\tilde{q}_k)\, .
  \label{parstcs:chis}
\end{eqnarray}

We remark that by setting $a=1$, as well as
$J_{2,0}=r_{2,0}=q_2=\tilde{r}_{2,0}=\tilde{q}_2=0$, $\tilde{q}_{2,0}=\tilde{q}_{1,0}=0$ and
$q_0=1$ we exactly recover the free energy for the Hopfield
model with synchronous updating and variable dilution \cite{tonithe}.

\subsection{Saddle-point equations}

The extremization of the free energy in (\ref{beforetracef}) yields
\begin{eqnarray}
&&m_k=\frac{1}{a_k}\left\langle\eta_k
     \left\langle\sigma^k\right\rangle_{\beta}\right\rangle_{\xi^c,z}\, ,
     \label{finalsaddle0} \\
&&q_0=\left\langle \left\langle
   \sigma^2\right\rangle_{\beta}\right\rangle_{\xi^c,z}\, ,\quad
  q_k=\left\langle \left\langle
      \sigma^k\right\rangle_{\beta}^2\right\rangle_{\xi^c,z}\, ,\\
&&r_{k,0}=\left\langle \left\langle
    \sigma^k\tau^k\right\rangle_{\beta}\right\rangle_{\xi^c,z}\, ,
    \label{finalsaddle1}
\end{eqnarray}
with $\left\langle \cdot \right\rangle_{z}$ the average over the Gaussian noise (recall the last term in (\ref{beforetracef})) and  $\left\langle \cdot \right\rangle_{\beta}$ the thermal average performed with respect to (\ref{beforetraceH}).
The conjugate variables satisfy
\begin{eqnarray}
&& \tilde{q}_{k}=\frac{q_k}{a_k^2}\left(1-c_k+\frac{c_k}{(1-\chi_{kr}-\chi_k)^2}\right)\, ,
           \label{eq:parstcs:qtil} \\
&& \tilde{\chi}_k=\frac{\chi_k}{a_k^2}
  \left(1-c_k+\frac{c_k}{(1-\chi_{kr})^2-\chi_k^2}\right)\, ,
         \label{eq:parstcs:chitil} \\
&& \tilde{\chi}_{kr}=\frac{1}{a_k^2}\left((1-c_k)\chi_{kr}+ \frac{c_k(1-\chi_{kr})}
               {(1-\chi_{kr})^2-\chi_k^2}+J_{k,0}-c_k\right)\, .
           \label{eq:parstcs:chirtil} \nonumber \\
\end{eqnarray}

In the absence  of thermal noise ($T= 0$), the system reaches a minimum of the free energy
such that $\sigma_i=\tau_i$, $\forall i$.  Therefore, it is interesting to consider the 
limit $\sigma=\tau$ of the free energy and the saddle-point equations. The trace of the effective partition function (recall (\ref{beforetraceH})) has to be evaluated again and, in addition, the following equivalences hold
\begin{eqnarray}
&& r_{k,0}=q_0\, ,\quad
\chi_{kr}=\chi_k\, ,\\
&&\tilde{\chi}_{kr}=\tilde{\chi}_k\, ,\quad
\tilde{r}_{k,0}=\tilde{q}_{k,0}+\frac{1}{a_k\beta}J_{k,0}\, .
\end{eqnarray}
For zero self-couplings ($J_{k,0}=0$), this leads to the relation $\beta f_{syn}(\beta)= 2\beta f_{seq}(2 \beta)$, as is also the case for the $Q=3$ Ising network \cite{tonithe}.
This is easy to understand when noticing that for $\tau_j^k=\sigma_j^k$ the  Hamiltonian
(\ref{hamtwospin}) for synchronous updating becomes simply twice the one for sequential updating. An immediate consequence is that the saddle-point equations are identical to those for sequential updating \cite{BV03,tonithe}.
 
For the saddle-point equations at $T=0$ we then obtain, with obvious notation
\begin{eqnarray}
&& m_k= \frac{1}{a_k}\left\langle\eta_k 
\textrm{sign}^k(\tilde{h}_1(z_1))\, {\Theta}(|\tilde{h}_1(z_1)|+\tilde{h}_2(z_2))
         \right\rangle_{\xi^c,z} \nonumber \\ \\
&& q_0 =\left\langle{\Theta}(|\tilde{h}_1(z_1)|+
         \tilde{h}_2(z_2))\right\rangle_{\xi^c,z} \\
&& \chi_k =\frac{1}{a_k\sqrt{\alpha_k \tilde{q}_k}} \nonumber \\
&& \hspace{0.6cm} \times \left\langle z_k\, \textrm{sign}^k(\tilde{h}_1(z_1))\,                      {\Theta}(|\tilde{h}_1(z_1)|+
                \tilde{h}_2(z_2))\right\rangle_{\xi^c,z}
\end{eqnarray}
where the effective fields are given by
\begin{eqnarray}
\tilde{h}_k(z_k)&=&\frac{1}{a_k}\eta_km_k+
    \delta_{k,2}\sum_{k'=1}^2\alpha_{k'}\left(a_{k'}\tilde{\chi}_{k'} 
       +    \frac{J_{k,0}}{2a_{k'}}\right) \nonumber \\
    &+& \sqrt{\alpha_k\tilde{q}_k}z_k \,.
\label{forgoth1}
\end{eqnarray}
The conjugated variables read
\begin{eqnarray}
&& \tilde{q}_k =\frac{q_k}{a_k^2}\left(1-c_k+\frac{c_k}{(1-\chi_k)^2}\right)\, ,
     \label{eq:seqstcs:qtilpr}        \\
&&\tilde{\chi}_k =\frac{\chi_k}{a_k^2}\left(1-c_k+\frac{c_k}{1-\chi_k}\right)\, .
      \label{eq:seqstcs:chitilpr}
\end{eqnarray}

\section{Thermodynamic and retrieval properties}
\label{4}

In this Section we study the thermodynamic and retrieval properties of the BEG network with variable dilution and synchronous updating by numerically solving the saddle-point equations (\ref{finalsaddle0})-(\ref{eq:parstcs:chirtil}) and compare the results with those for sequential updating.
As in the model with sequential updating \cite{BV03}, there are four phases: the retrieval phase $R$ ($m_k>0$, $q_1>0$), the spin-glass phase $SG$ ($m_k=0$, $q_1>0$), the paramagnetic phase $P$ ($m_k=0$, $q_1=0$) and the quadrupolar phase $Q$ ($m_1=0$, $m_2>0$, $q_1>0$). 
In the sequel we discuss the phase diagrams for low loading ($ \alpha_k=0$), finite loading 
($ \alpha_k>0$), $k$-type dilution and the effect of the self-couplings $J_{k,0}$.

\begin{figure}[ht]
\begin{center}
\resizebox{0.8\columnwidth}{!}{
 \includegraphics*[angle=270,scale=0.65]{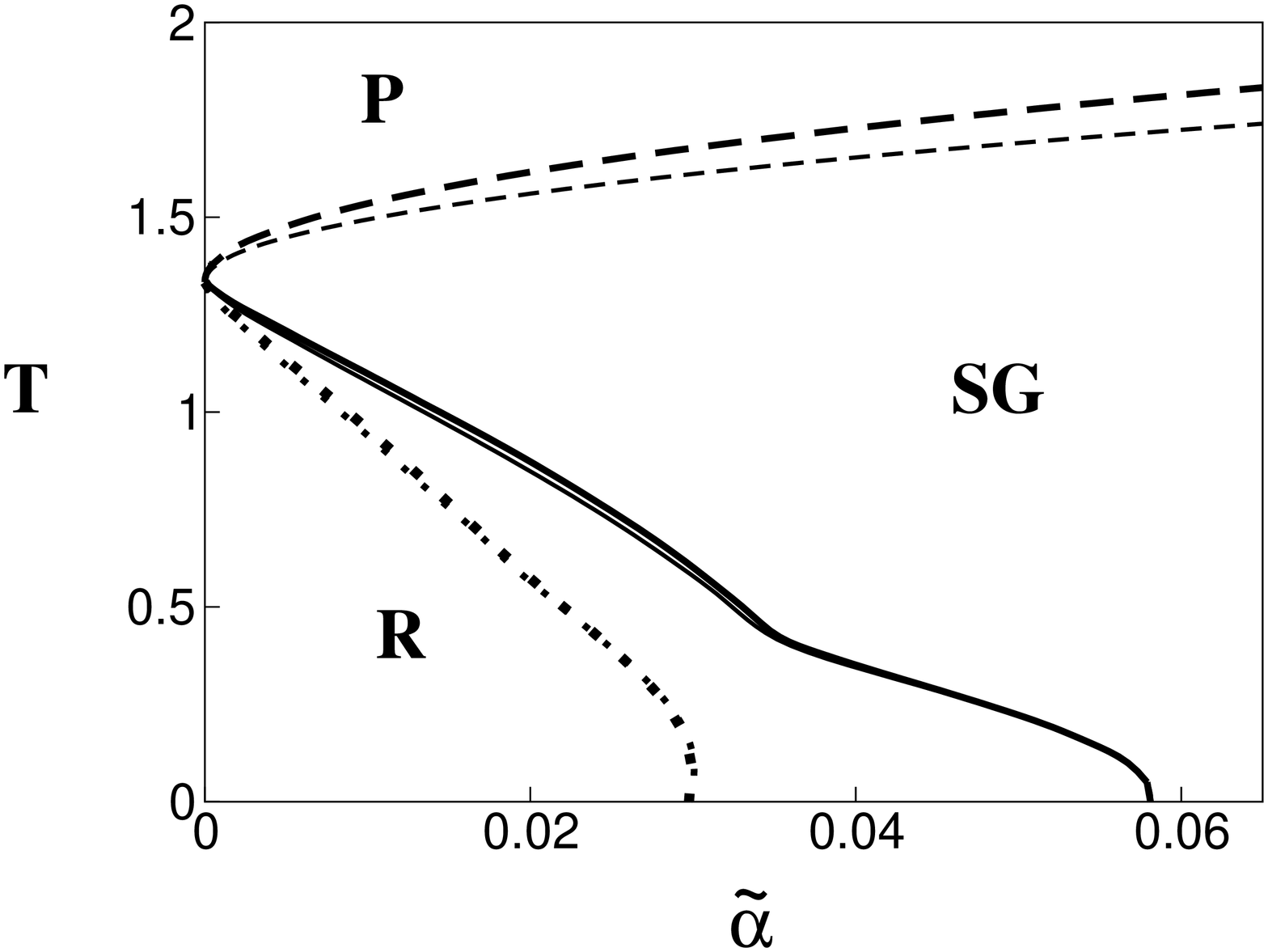}}
 \\
\resizebox{0.8\columnwidth}{!}{ 
 \includegraphics*[angle=270,scale=0.65]{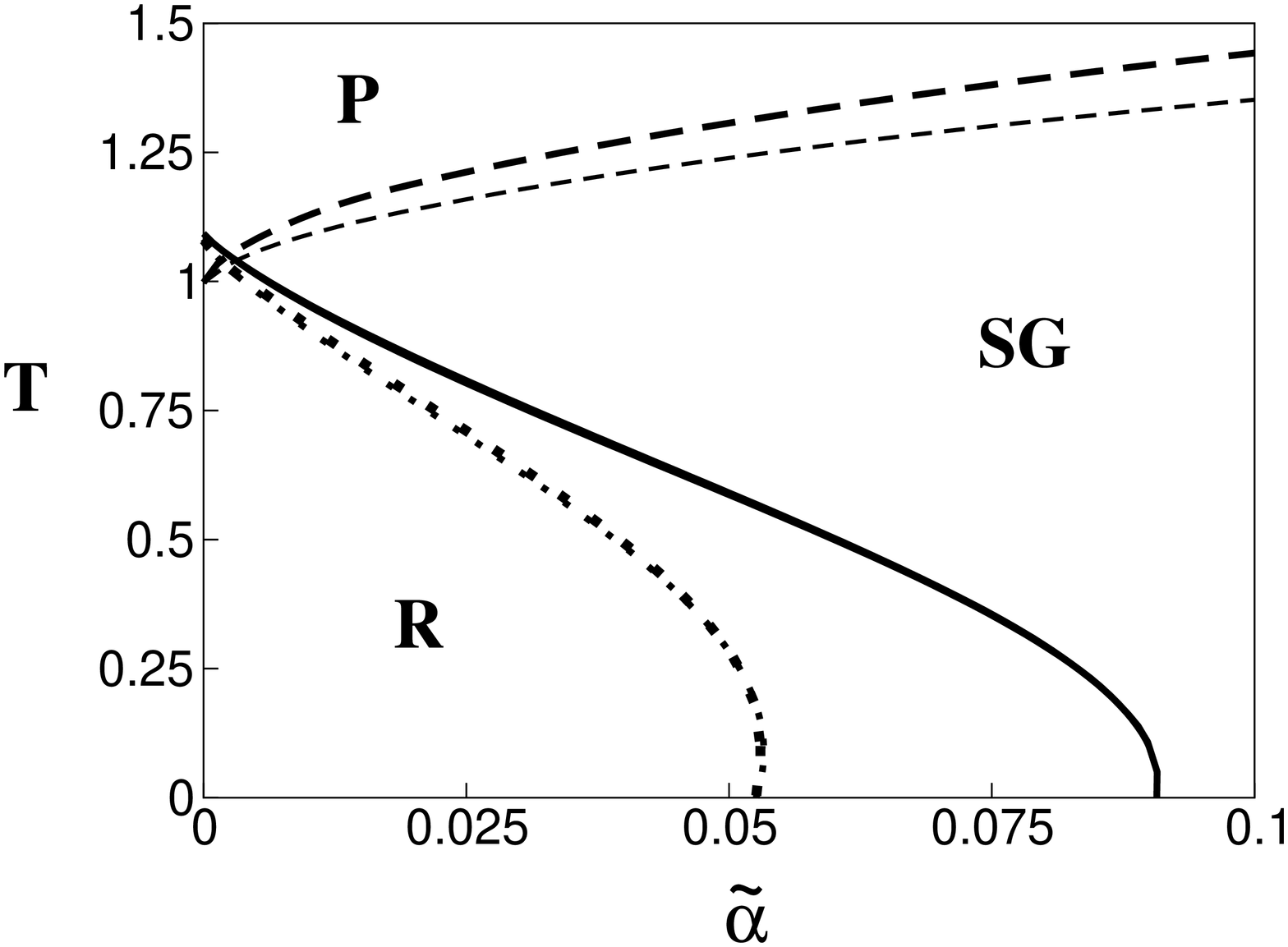}}
\\
\resizebox{0.8\columnwidth}{!}{ 
 \includegraphics*[angle=270,scale=0.65]{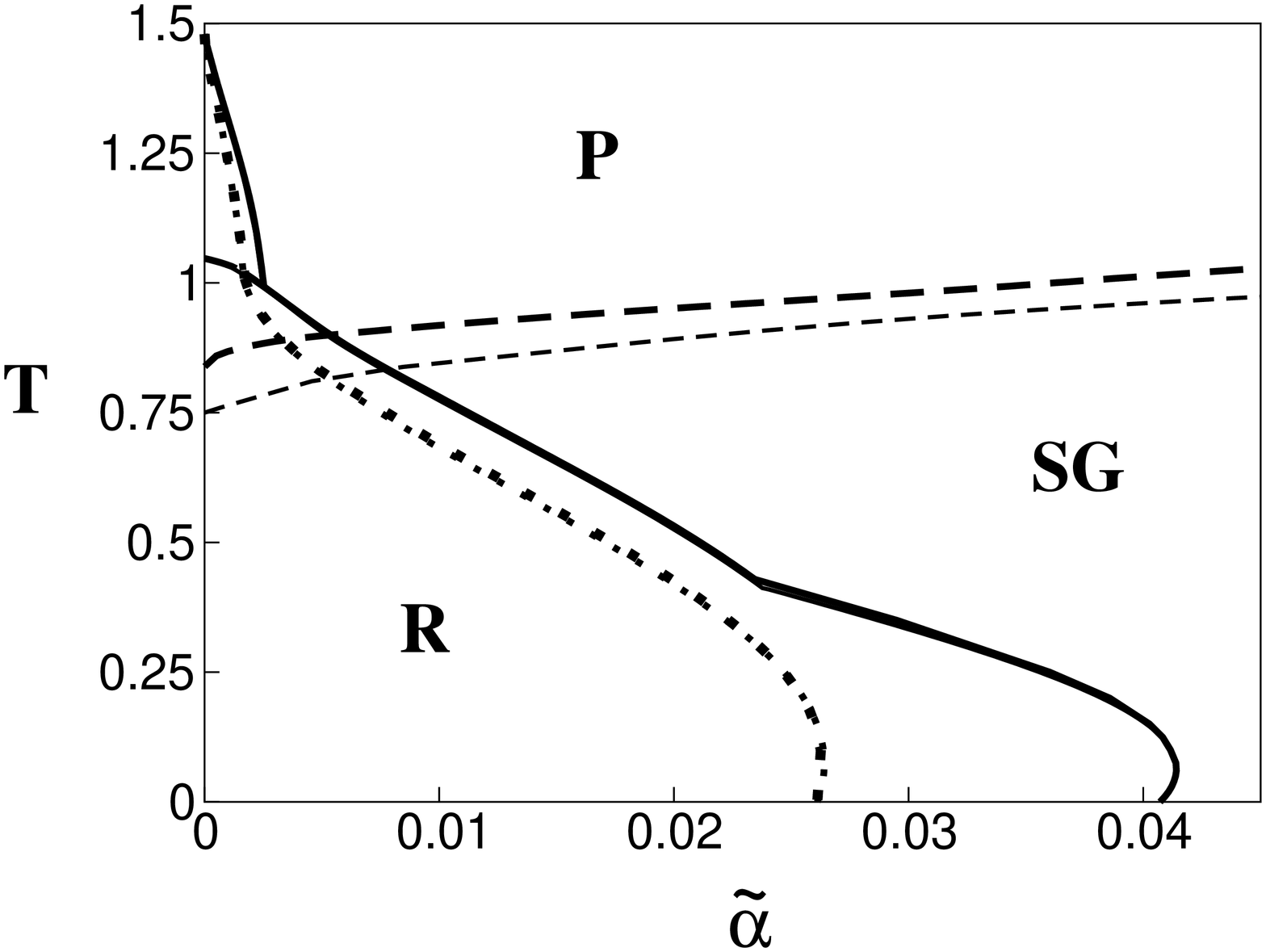}}
\caption{The $\tilde{\alpha}-T$ phase diagram for the fully connected  BEG neural network with activities $a=0.5$ ,$2/3$ and $0.8$ for synchronous and
sequential updating (thick respectively thin lines). }
   \label{fig:parstcs:pd_par_vs_seq}       
\end{center}
\end{figure}
\begin{figure}[ht]
\begin{center}
\resizebox{0.8\columnwidth}{!}{
 \includegraphics*[angle=270,scale=0.65]{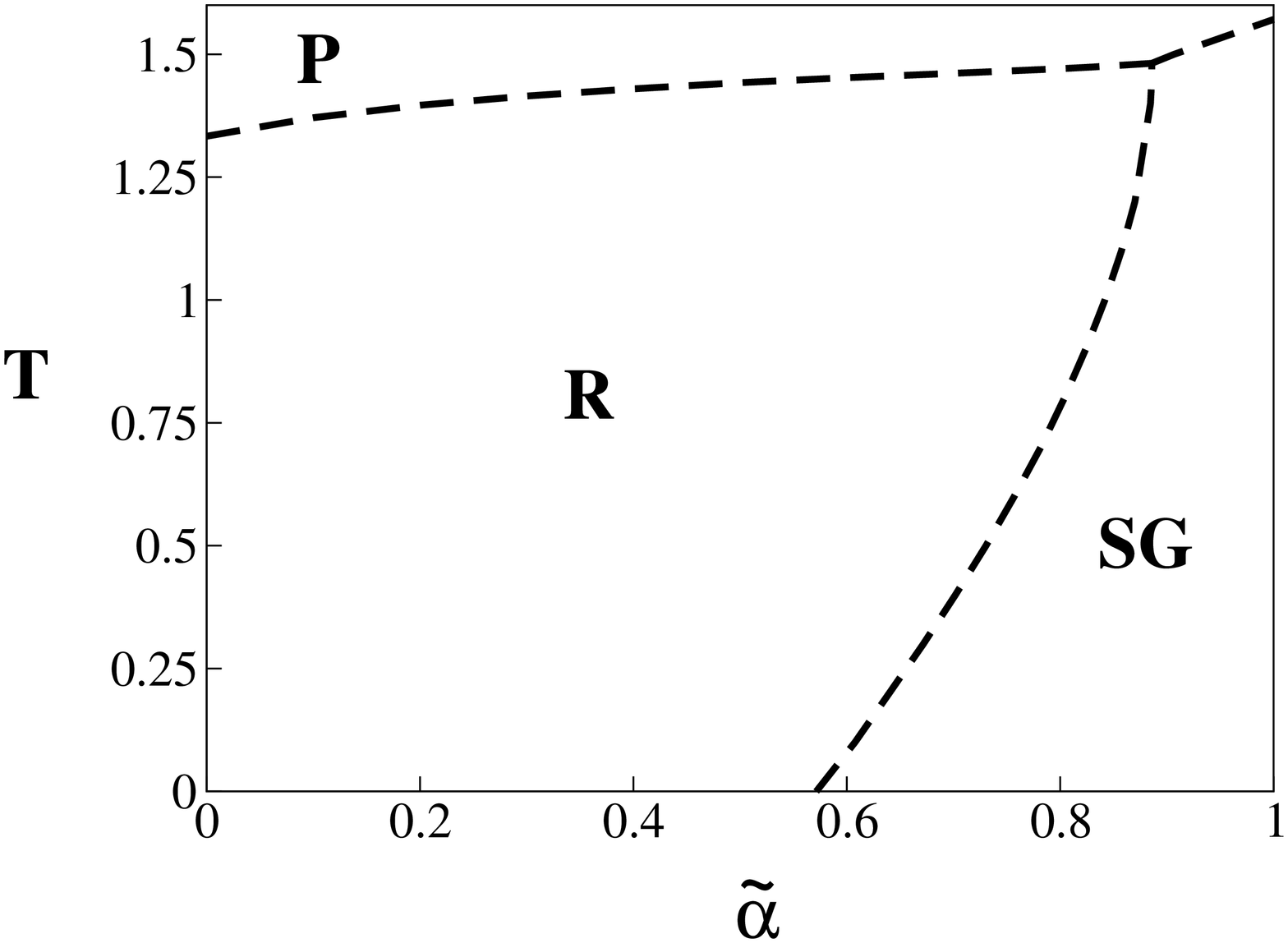}}
 \\
\resizebox{0.8\columnwidth}{!}{ 
 \includegraphics*[angle=270,scale=0.65]{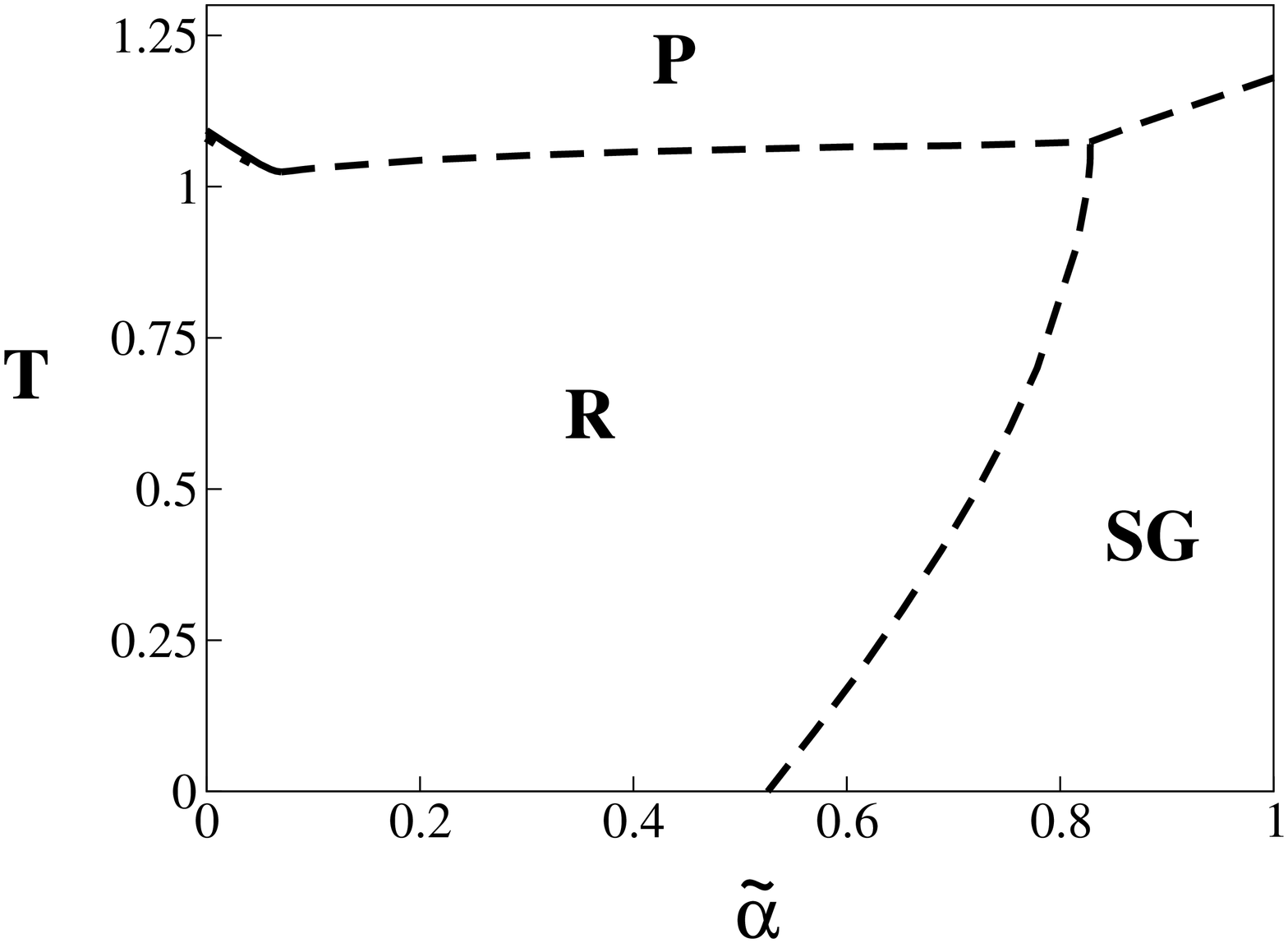}}
\\
\resizebox{0.8\columnwidth}{!}{ 
 \includegraphics*[angle=270,scale=0.65]{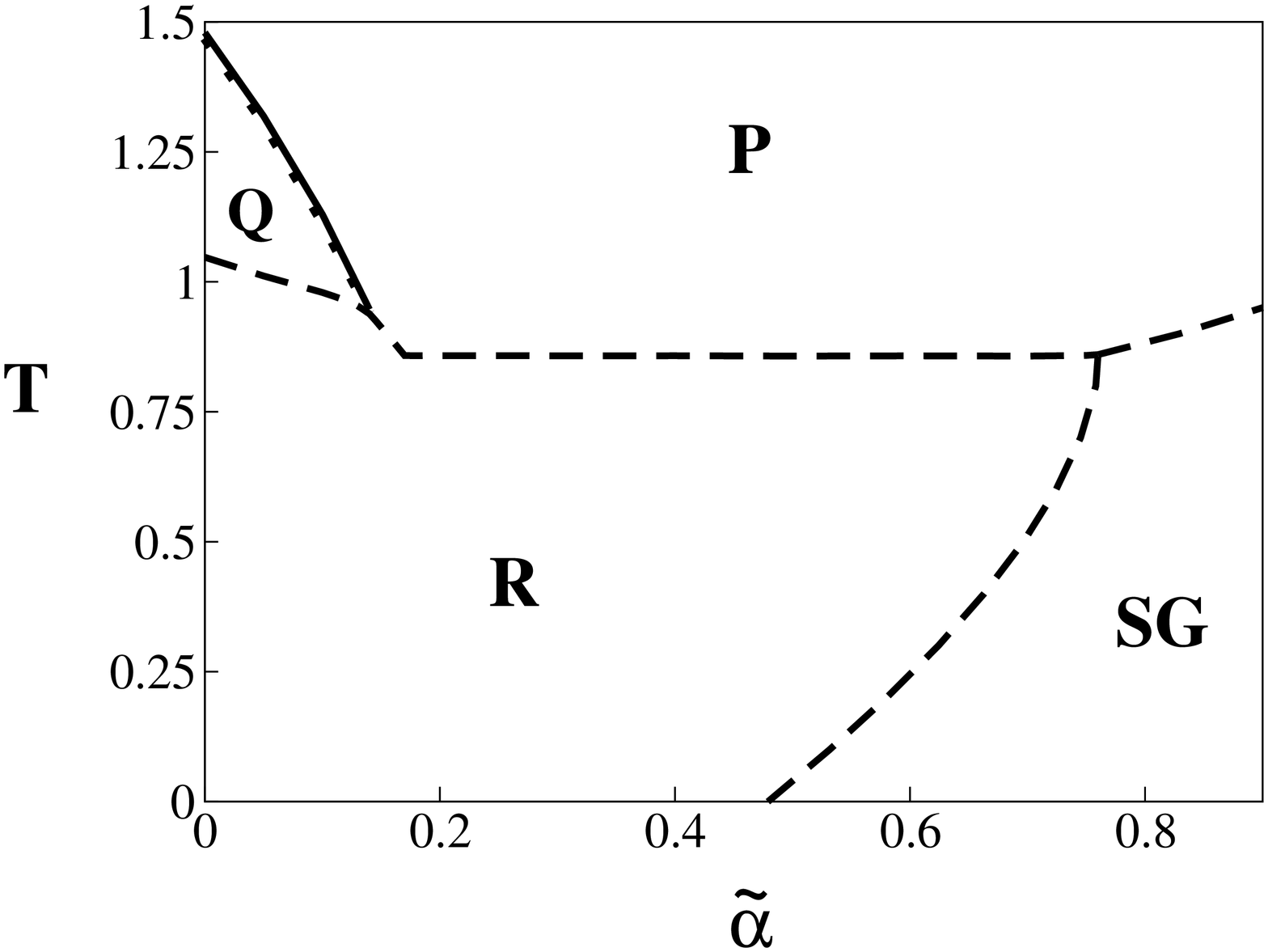}}
\caption{The $\tilde{\alpha}-T$ phase diagram for the extremely diluted BEG neural network with activities $a=0.5$,  $2/3$ and $0.8$. }
   \label{fig:parstcs:ed_seq}       
\end{center}
\end{figure}

\subsection{Low loading}

For $ \alpha_k=0$, the saddle-point equations can  be derived without the use of replicas (e.g., \cite{AGS85,vH86} for the Hopfield model) and we find 
\begin{equation}
\label{mpar}
\hspace{-0.117cm} m_{k,\sigma}=
\frac{1}{a_k}\left\langle\eta_k\frac{\underset{\sigma}{\trace}\,\sigma^k\exp{\left(
\beta\sum_{k'}\frac{1}{a_k}\eta_{k'} m_{k',\tau}\sigma^{k'}\right)}}
{\underset{\sigma}{\trace}\,\exp{\left(
\beta\sum_{k'}\frac{1}{a_k}\eta_{k'} m_{k',\tau}\sigma^{k'}\right)}}\right\rangle_{\xi^c}
     \\
\end{equation}
and similarly for $m_{k,\tau}$.  One can show that the only minimum of the free energy is given by $m_{k,\sigma}=m_{k,\tau}$, such that the solution coincides which the one for sequential updating. Physically, this means that in equilibrium there can be no cycles: the two neuron variables have decoupled becoming independent and leading to $\beta f_{syn}(\beta)= 2\beta f_{seq}(\beta)$, as seen in the other models mentioned above \cite{BB04,tonithe,AGS85,vH86}.
Hence, the dynamics does not influence the equilibrium solutions and the $a-T$ phase diagram is identical to the one for sequential updating. Furthermore, we know \cite{BV03} that for low loading the form of the architecture (fully connected, extremely diluted) is not important. Hence we do not treat this case in more detail here but we refer to, e.g., \cite{BV03} (Figure 1) and the discussion therein.

\subsection{Finite loading}

For the fully connected architecture, $c_k=1$, Fig.~\ref{fig:parstcs:pd_par_vs_seq} presents the capacity-temperature phase diagrams for three typical values of the activity
$a=0.5$, $2/3$ (uniformly distributed patterns) and $0.8$. 
For the moment, the self-couplings are set to zero and the dilution parameters are equal, i.e., $c_1=c_2$ (recall that in that case $\alpha_k= \tilde{\alpha}$).  Dashed lines correspond to second
order transitions, full lines to first order transitions and  dotted lines to thermodynamic transitions. The thick lines indicate the results for synchronous updating;
for comparison we have indicated those for sequential updating \cite{BV03} with thin lines.

As can be seen the main features for both types of updating are very similar. For synchronous updating the spin glass region increases  visibly while the retrieval region increases marginally. A similar behaviour is observed for the Hopfield model \cite{FK88,tonithe}. 
Finally,  for $a=0.8$,  the quadrupolar region is slightly enlarged as well.

For extreme symmetric dilution ($c_k=0$) the phase diagrams nearly coincide with the ones for sequential updating \cite{tonithe}. Since both sets of results did not yet appear in the literature we show them in Fig.~\ref{fig:parstcs:ed_seq}. 

Compared with the fully connected architecture, most of the first order transitions become second order. Only the quadrupolar-paramagnet first order transition remains. Furthermore, there is strong re-entrant behaviour in the retrieval region. This is related to the RS approximation and is also seen in the Hopfield model \cite{WS91,CN92} and the $Q=3$-Ising model \cite{BCS00}.

\subsection{${\bf k}$-type dilution}

Next, we look at the effect of $k$-type dilution on the system (recently introduced as asynchronous dilution  in a Gardner optimal capacity calculation \cite{BC05}). Indeed, the BEG model allows one to prune independently the two types of couplings appearing in the Hamiltonian (\ref{hamtwospin}).


The results can be summarized as follows. The performance of the network strongly depends on the activity. In any case, the {\it absolute}  capacity $p/N$ decreases when the pruning becomes stronger, as one expects, while the {\it standard}  capacity $p/cN$ shows a maximum. 

\begin{figure}[ht]
\begin{center}
\resizebox{0.8\columnwidth}{!}{
 \includegraphics*[angle=270,scale=0.65]{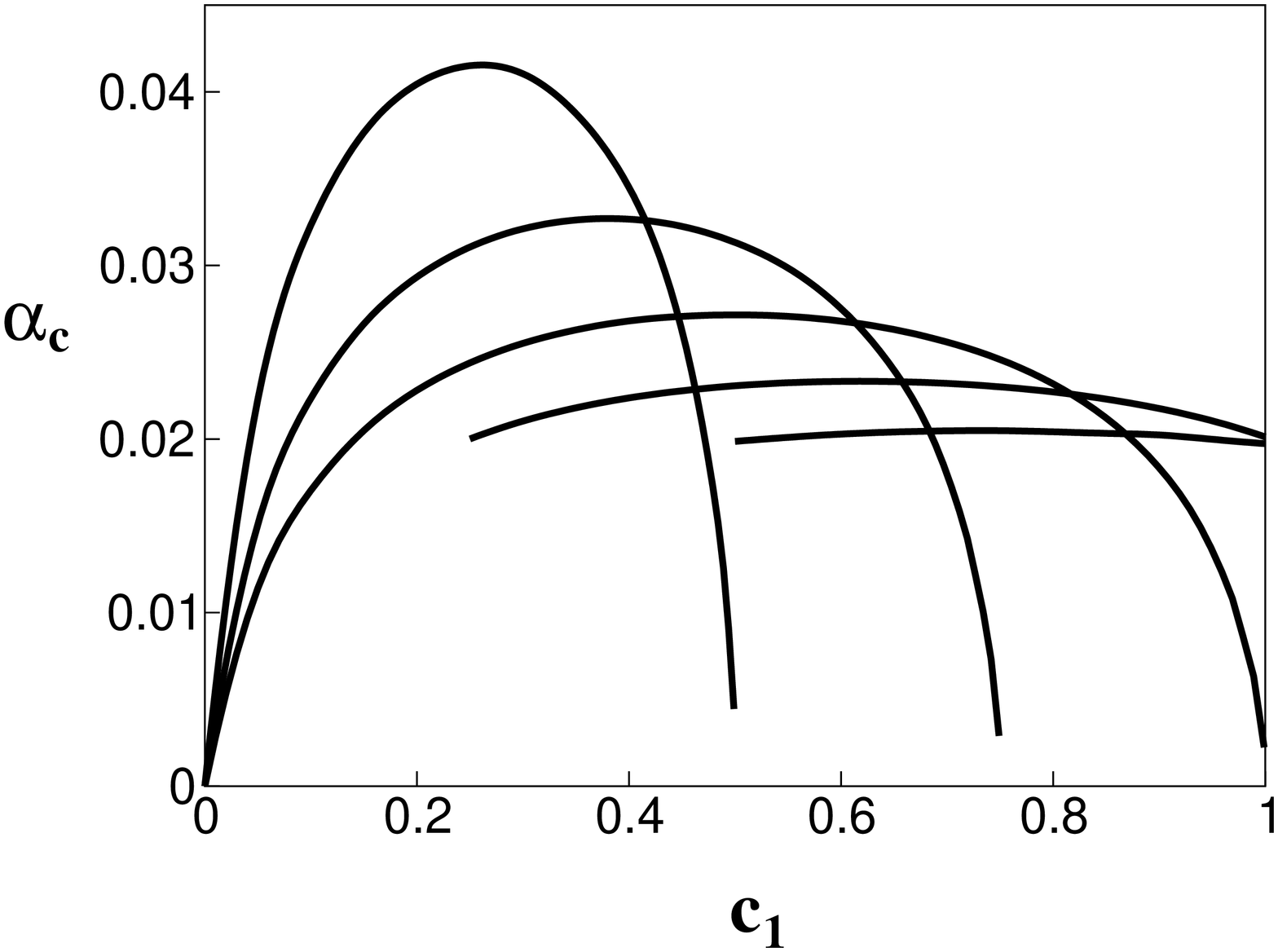}}
 \\
\resizebox{0.8\columnwidth}{!}{ 
 \includegraphics*[angle=270,scale=0.65]{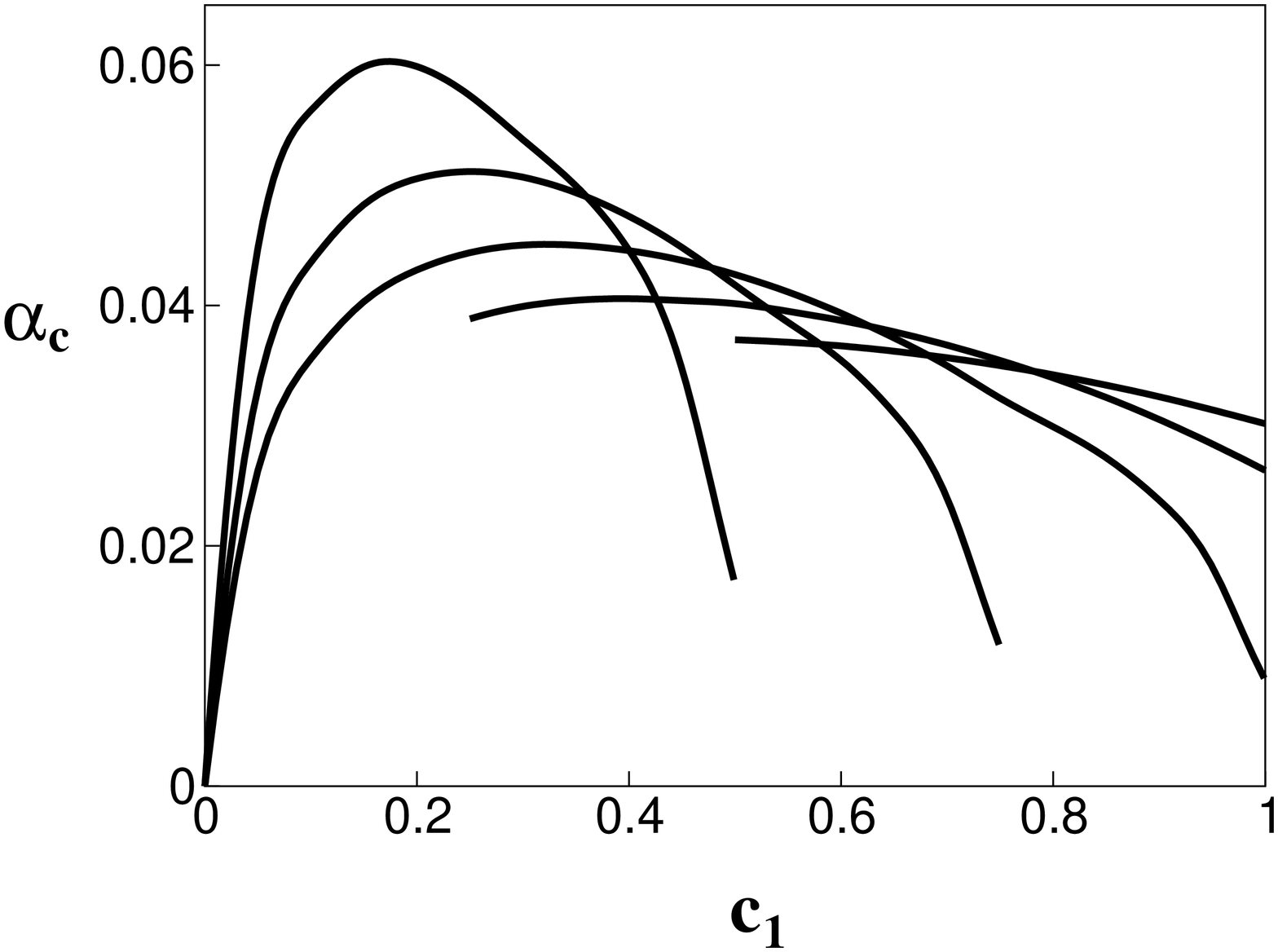}}
\\
\resizebox{0.8\columnwidth}{!}{ 
 \includegraphics*[angle=270,scale=0.65]{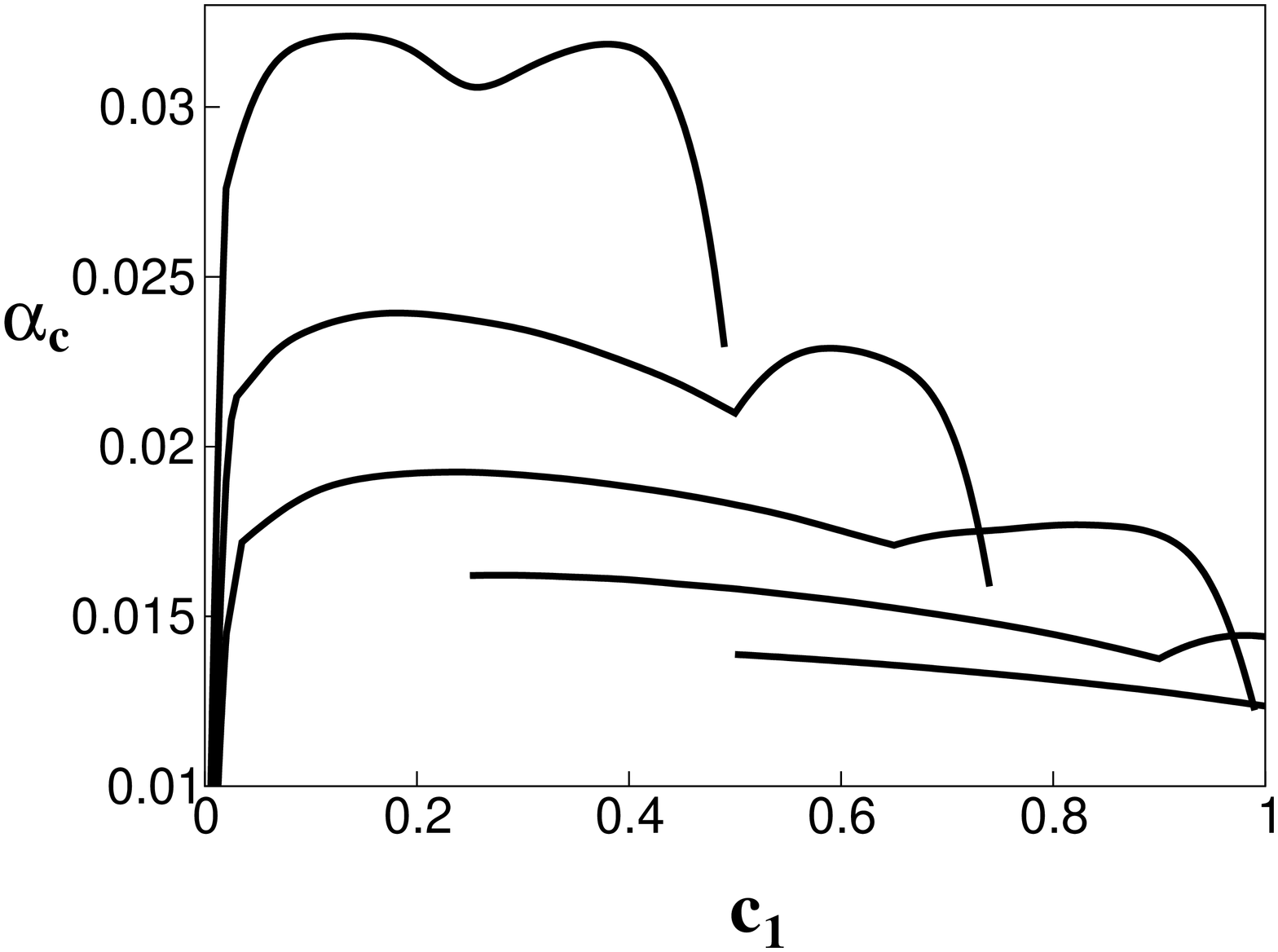}}
\caption{Standard critical capacity at $T=0.5$ as a function of $c_1$. From top to bottom,    $a=0.5$, $2/3$ and $0.8$. In each figure, from top to bottom,  $c=0.5$, $0.75$, $1.0$, $1.25$ and $1.5$.  }
   \label{c_fixed}       
\end{center}
\end{figure}

Hence, given the number of connections per neuron, $c= c_1+c_2$, there has to be
an optimal combination of both dilution variables $c_1$ and $c_2$. In order to be able to predict this combination once  the other parameters of the system are known we have performed some numerical experiments. We present in Fig.~\ref{c_fixed} the critical capacity $\alpha_c$ for the activities $a=0.5$, $2/3$ and $0.8$ as a function of $c_1$. The values for $c$ are $c=0.5$, $0.75$, $1.0$, $1.25$ and $1.5$, from top to bottom, and the temperature is fixed at $T=0.5$. 
We find a maximum in all figures, corresponding to the optimal combination of the dilution parameters $c=c_1^*+c_2^*$. 
Looking, e.g., at the optimal value $c_1^*$ as a function of the number of connections per neuron $c$ for $T=0.5$ for several values of the activity, $a=0.9$, $0.8$, $2/3$, $0.5$ and $0.4$ we find that these data points can be fitted by
\begin{equation}
c_1^*=(1-a)(c_1+c_2)=(1-a)c\, .
\end{equation} 
We conjecture that the temperature does not modify the position of the maximum but only its height, since $c_1^*$ is determined by the role of both terms in the Hamiltonian and therefore by the activity. 

Furthermore, we notice a second maximum for $a=0.8$. This is due to the existence of a second retrieval solution occurring for large activities \cite{BV03}. This solution only appears for $0< c \lesssim 1.4$ (for $T=0.5$). We can fit this second peak by 
\begin{equation}
c_1^*=a(c_1+c_2)=ac\, .
\end{equation}
Since for larger activities this relation implies that there are fewer $k=2$-type connections at the maximum $c_1^*$, the second retrieval solution has a smaller overlap $m_2$.

Furthermore, when $c_1$ is kept finite, both critical capacities for $c_2=0$ are always finite. However, when $c_2$ is kept finite, setting $c_1=0$ destroys any retrieval.
This suggests that it is actually the first term of the Hamiltonian that mainly ensures the process of retrieval. Keeping only the second term of the Hamiltonian, the system  still recognizes correctly the zero and the active states, but does not discern between the $+1$ and $-1$ states. It still implies though that the information content of the system is nonzero. A practical realization of this behaviour may be in pattern recognition where, looking at black and white pictures on a grey background, the system would tell us the exact location of the picture with respect to the background without finding the details of the picture itself.  These findings seem to be in agreement with the results in \cite{BC05} for the Gardner optimal capacity in the presence of dilution, where it is concluded that the BEG network is indeed more robust against $k=2$-type dilution of the couplings.

\subsection{The self-couplings effect}

Finally, we turn to the study of the effect of self-coupling on the phase diagrams of the model. In the case of very large but finite self-couplings $J_{1,0}$ and/or $J_{2,0}$, the $\sigma$ and $\tau$ tend to stay in the same state. Then, the Hamiltonian becomes
\begin{eqnarray}
&&\lim_{J_{k,0}>>1}H_{syn}(\bsigma) \nonumber \\
&& \hspace{0.5cm}=2H_{seq}(\bsigma)-
  \sum_{k=1}^2\frac{\alpha_k}{a_k}J_{k,0}\sum_{i=1}^N\delta_{|\sigma_i|,1}\,\, .
\end{eqnarray}
The last term favours the elimination of the zero states from the system as can also be seen from eqs.(\ref{three})-(\ref{four}) in these limits. The order parameters of the system evolve to  $m_2=0$, $q_0=1$, $q_2=1$, $r_{2,0}=1$ and   
\begin{description}

\item (i) for $J_{1,0}\rightarrow\infty$ or  $J_{1,0}$ and  $J_{2,0}\rightarrow\infty$  

\begin{eqnarray}
m_1&=&\frac{1}{a_1}\left\langle\eta_1\tanh{(2\beta\tilde{h}_1(z_1))}
     \right\rangle_{\xi^c, z_1}  \\
q_1&=&\left\langle\tanh^2{(2\beta\tilde{h}_1(z_1))}
    \right\rangle_{\xi^c, z_1}\\ 
r_{1,0}&=&1.
\end{eqnarray}
\item (ii) for $J_{2,0}\rightarrow\infty$

\begin{eqnarray}
m_1&=&\frac{1}{a_1}\left\langle \eta_1\frac{\sinh{(2\beta\tilde{h}_1(z_1))}}
{\cosh{(2\beta\tilde{h}_1(z_1))}+e^{-2\beta\alpha_1 a_1\tilde{\chi}_{1r}}}\right\rangle_{\xi^c,z_1} \nonumber \\ \\
q_1&=&\left\langle\frac{\sinh^2{(2\beta\tilde{h}_1(z_1)}}
{\cosh^2{(2\beta\tilde{h}_1(z_1))}+e^{-4\beta\alpha_1 a_1\tilde{\chi}_{1r}}}\right\rangle_{\xi^c,z_1} \\
r_{1,0}&=&\left\langle\frac{\cosh{(2\beta\tilde{h}_1(z_1))}-e^{-2\beta\alpha_1a_1\tilde{\chi}_{1r}}}
{\cosh{(2\beta\tilde{h}_1(z_1))}+e^{-2\beta\alpha_1 a_1\tilde{\chi}_{1r}}}\right\rangle_{\xi^c,z_1}
\end{eqnarray}
\end{description}
with $\tilde{h}_1(z_1)$ given by (\ref{forgoth1}) ($k=1$). As can be seen from these equations, in the first limiting case we arrive at a Hopfield model with sequential updating when $a\rightarrow 1$,  with the usual rescaling in $\beta$ ($\beta \rightarrow \beta/2 $).  
For the second limiting case, we only evolve to the Hopfield model with sequential updating in the  limit $\beta\rightarrow\infty, a \rightarrow 1$. Indeed, in the first case, the second local field $h_{2}$ becomes irrelevant when $a=1$,  and the Hamiltonian reduces to a Hopfield-like one with a rescaled $\beta$ and $\sigma_i=\tau_i$ for any temperature. In the second case
$\sigma_i^2=\tau_i^2$ for any temperature, but only for $T=0$ we are ensured that $\sigma_i=\tau_i$ as well.

\begin{figure}[ht]
\begin{center}
\resizebox{0.8\columnwidth}{!}{
 \includegraphics*[angle=270,scale=0.65]{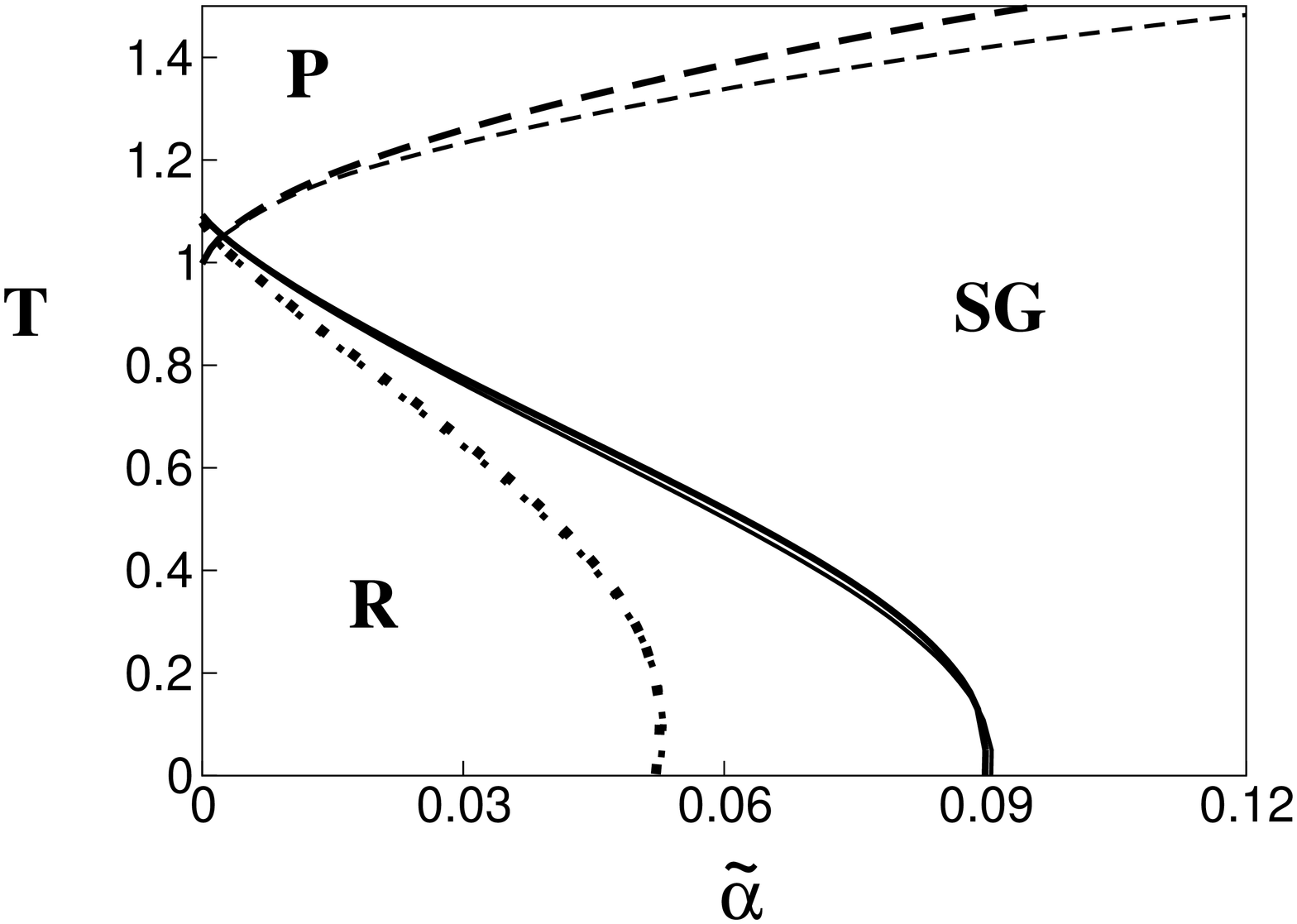}}
 \\
\resizebox{0.8\columnwidth}{!}{ 
 \includegraphics*[angle=270,scale=0.65]{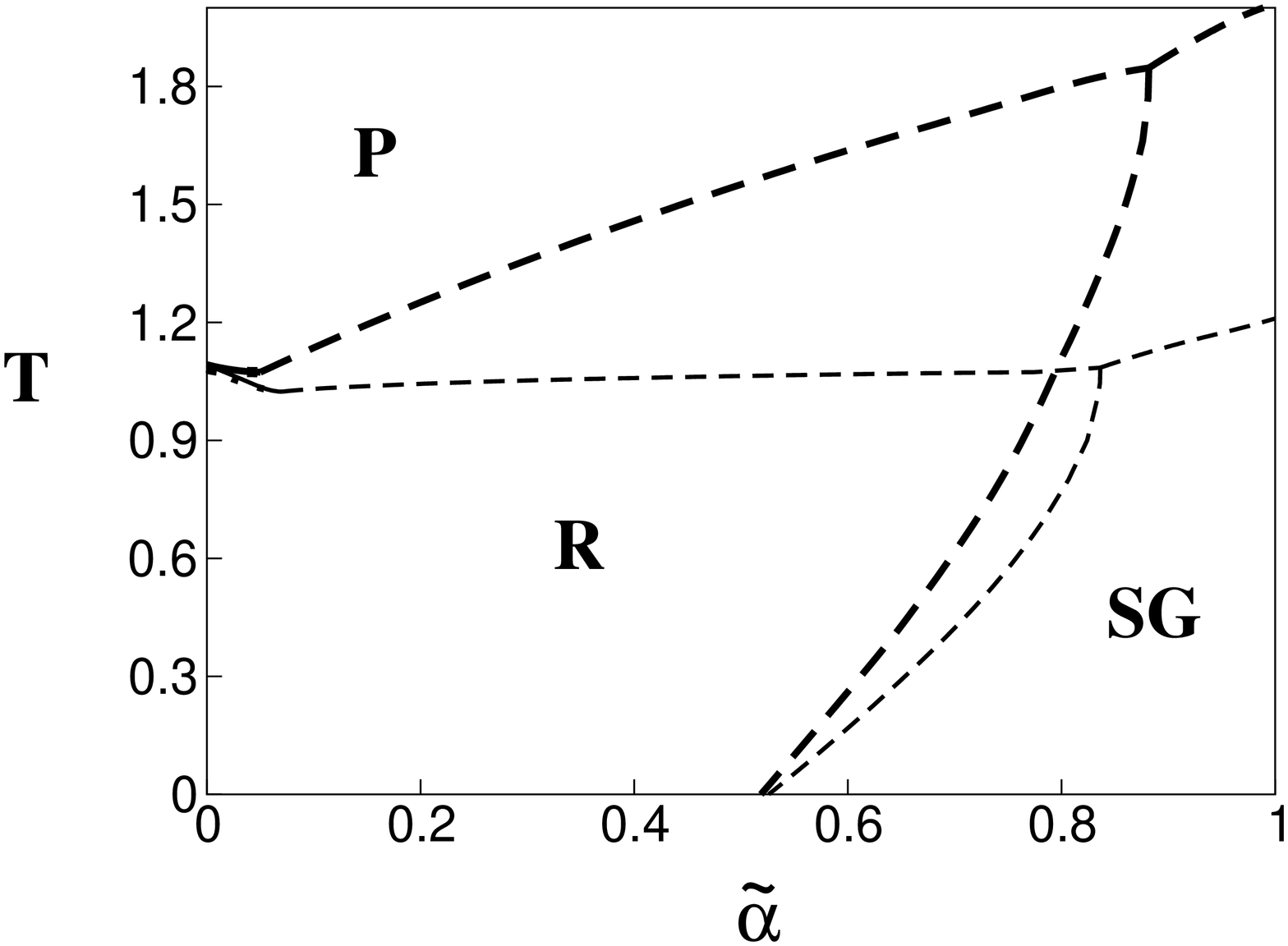}}
\caption{ The $\tilde{\alpha}-T$ phase diagrams for $a=2/3$ with
      $J_{k,0}=0$ (thin lines) and $J_{k,0}=a_k$ (thick lines) for the fully connected 
      (top) and the extremely diluted (bottom) BEG networks. }
   \label{pd_unif_pat_par_self_coup}       
\end{center}
\end{figure}

In Fig.~\ref{pd_unif_pat_par_self_coup} some phase diagrams in the presence of self-couplings $J_{k,0}=a_k$ are shown in comparison with the results without self-couplings. 
For the fully connected architecture, $c_k=1$, there is little  difference between synchronous and sequential updating. While the spin glass region is visibly enlarged, the retrieval region has increased only marginally. 
  
The extremely diluted network is much more sensitive to the presence of self-couplings. The retrieval region is substantially enlarged, as well as the spin glass region.
A new feature is the different re-entrance point for different values of the self-coupling. This is not observed in the Hopfield model \cite{tonithe} because the limit of large self-couplings in that case is the Hopfield model for sequential updating, which is known to have the same RS critical capacity. Here the phase diagrams must evolve from the BEG phase diagram with synchronous updating to the Hopfield one with sequential updating, which has a different critical capacity. Therefore, the re-entrance point also evolves as we increase $J_{k,0}$. A similar effect has been observed for the $Q$-Ising model with finite
dilution ($c=0.01$, \cite{tonithe}).

\begin{figure}[ht]
\begin{center}
\resizebox{0.8\columnwidth}{!}{
 \includegraphics*[angle=270,scale=0.65]{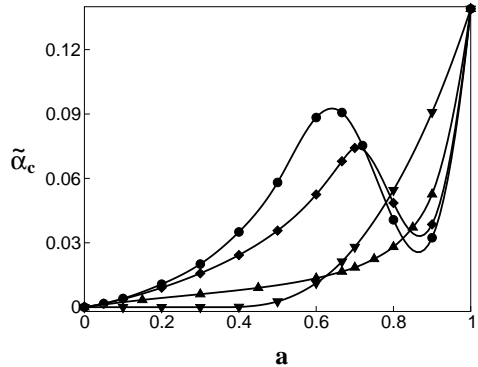}}
  \caption{Critical capacity as a function of the activity $a$ for $T=0$ and
different values of the self-couplings $J_{k,0}=0$ (circles), 10 (diamonds), 100 (triangles up) and $\infty$ (triangles down).}
        \label{alpha_c}
\end{center}
\end{figure}

Hence, the self-couplings in three-state neural networks with variable dilution are very relevant. Different values can produce families of phase diagrams that evolve towards the Hopfield phase diagram for sequential updating with the rescaling in $\beta$ mentioned before. An illustration of this effect is presented, for $T=0$, in Fig.~\ref{alpha_c}, where the critical capacity is shown for different values of 
the self-couplings as a function of the activity. 
We remark that for $a=1$ we recover the value for the Hopfield model, $\alpha_c\simeq 0.139$. A maximum in the critical capacity appears unless the self-couplings are large enough so that the three-state nature of the model is breaking down since all zero states  are completely eliminated.

\section{Concluding remarks}
\label{6}

We have considered the thermodynamic and retrieval properties of BEG networks with synchronous updating and variable dilution.  Saddle-point equations for the relevant order parameters have been derived for arbitrary temperature using a replica mean-field treatment. In the stationary limit, the system allows for fixed-points as well as two-cycles. It is found numerically that two-cycles only occur in the spin-glass phase of the system, never involve a large number of neurons and are always close to a fixed-point spin glass solution. 

Making a non-cycle and replica symmetry ansatz, the phase diagrams of the model are obtained. Two types of dilution involving the two types of couplings are discussed. The optimal combination of the network dilution parameters giving the largest critical capacity is obtained. It is also found that the model is more robust against dilution of the $k=2$-type couplings. 

The results for different pattern activities and variable dilution are compared with those for sequential updating. It is observed that the asymptotic behaviour of the network with synchronous and sequential updating is almost identical. For some values of the network parameters the retrieval region is enhanced marginally. Only the presence of self-couplings can enlarge the retrieval region substantially, especially in the case of extreme dilution.

\section*{Acknowledgements}

The authors would like to thank  Rubem Erichsen Jr., Isaac P\'erez Castillo, Gyoung Moo Shim, Nikos Skantzos and Toni Verbeiren for useful discussions.
This work has been supported in part by the Fund of Scientific Research, Flanders-Belgium.


\begin{thebibliography}{99}
\bibitem{BB04}
D. Boll\'e and J. Busquets Blanco, Eur. Phys. J. B {\bf 42}, 397 (2004).
\bibitem{Pbook}
P. Peretto, {\it An introduction to the modeling of neural networks}, (Cambridge University Press: Cambridge, 1992), Chapter 3.
\bibitem{C01s} A.C.C Coolen, 
in \textit{Handbook of Biological Physics, 
    Vol.4} ed. F. Moss and S. Gielen, Elsevier Science B.V., 2001, p. 531.
\bibitem{FK88} 
J.F. Fontanari and R. K\"oberle, J. Phys. France {\bf 49}, 13 (1988);
J.F. Fontanari and R. K\"oberle, Phys. Rev. A {\bf 36}, 2475 (1987).
\bibitem{DK00}
  D.R.C. Dominguez and E. Korutcheva, Phys. Rev E {\bf 62}, 2620 (2000).
\bibitem{BV02}
  D. Boll\'e and T. Verbeiren, Phys. Lett. A {\bf 297}, 156 (2002).
\bibitem{bollebook}
D. Boll\'e, 
in \textit{Advances in condensed matter and statistical
mechanics}, eds. E. Korutcheva and R. Cuerno, Nova Science Publishers (New York,
2004), p. 319.
\bibitem{BV03} 
  D. Boll\'e and T. Verbeiren, J. Phys. A {\bf 36}, 295 (2003).
\bibitem{BDEKT03}
  D. Boll\'e, D.R.C. Dominguez, R. Erichsen Jr., E. Korutcheva and W.K. Theumann,
  Phys. Rev. E {\bf 68}, 062901 (2003).
\bibitem{tonithe} 
  T. Verbeiren, ``Dilution in recurrent neural
  networks'', Ph.D. thesis, K.U.Leuven, Leuven, Belgium 2003.
\bibitem{Peretto} 
  P. Peretto, Biol. Cybern. {\bf 50}, 51 (1984).
\bibitem{WS91} 
  T.L.H. Watkin and D. Sherrington, Europhys.\ Lett. {\bf 14}, 791 (1991)	
\bibitem{CN92}
  A. Canning and J.-P. Naef, J.\ Phys.\ I France {\bf 2}, 1791 (1992).     
\bibitem{VB85} 
  L. Viana and A.J. Bray, J.\ Phys.\ C: Solid State Phys.\  {\bf 18}, 3037 (1985).
\bibitem{BCS76} 
  R. Bidaux, J.P. Carton and G. Sarma, J.\ Phys.\ A {\bf 9}, L87 (1976)	              
\bibitem{TE01} 
  W.K. Theumann and R. Erichsen Jr.,  Phys. Rev. E {\bf 64}, 061902 (2001).
\bibitem{jordithe} 
  J. Busquets Blanco, ``Statistical mechanics of the Blume-Emery-Griffiths neural
  network'', Ph.D. thesis, K.U.Leuven, Leuven, Belgium 2005. 
\bibitem{L74} 
  W.A. Little, Math. Biosci. {\bf 19}, 101 (1974).
\bibitem{H82} 
  J.J. Hopfield, Proc. Nat. Acad. Sci. USA {\bf 79}, 2554 (1982).
\bibitem{SMK96} 
  T. Stiefvater, K.-R. M\"uller and R. K\"uhn, Physica A {\bf 232}, 61 (1996).
\bibitem{AGS85}
  D.J. Amit, H.G. Gutfreund and H. Sompolinsky,  Phys. Rev. A {\bf 32}, 1007 (1985).
\bibitem{vH86} 
  J.L. van Hemmen,  Phys. Rev. A {\bf 34}, 3435 (1986).
\bibitem{BCS00}
 D. Boll\'e, D.M. Carlucci and G.M. Shim, J. Phys. A {\bf 33}, 6481 (2000).
\bibitem{BC05}
 D. Boll\'e and I. P\'erez Castillo, Physica A {\bf 349}, 548 (2005).
\end{thebibliography}
\end{document}